\documentclass{aa}
\usepackage{graphicx}
\usepackage{aalongtable}
\begin{document}

\newcommand{\arcm}{$^\prime$}
\newcommand{\arcs}{$^{\prime\prime}$}
\newcommand{\m}{$^{\rm m}\!\!.$}
\newcommand{\D}{$^{\rm d}\!\!.$}
\newcommand{\F}{$^{\rm P}\!\!.$}
\newcommand{\kms}{km~s$^{-1}$}
\newcommand{\ks}{km~s$^{-1}$}
\newcommand{\ms}{M$_{\odot}$}
\newcommand{\rs}{R$_{\odot}$}
\newcommand{\oc}{$O\!-\!C$}
\newcommand{\ubv}{\hbox{$U\!B{}V$}}
\newcommand{\bv}{\hbox{$B\!-\!V$}}
\newcommand{\ub}{\hbox{$U\!-\!B$}}

\title{Discovery, photometry, and astrometry of 49 classical nova
       candidates in M81 galaxy\thanks{Partly based on observations obtained at the Gemini Observatory
(acquired through the Gemini Science Archive), which is operated by the Association of Universities for
Research in Astronomy, Inc., under a cooperative agreement with the NSF on behalf of the Gemini
partnership: the National Science Foundation (United States), the Particle Physics and Astronomy Research
Council (United Kingdom), the National Research Council (Canada), CONICYT (Chile), the Australian Research
Council (Australia), CNPq (Brazil) and CONICET (Argentina)}}

\author{  K. Hornoch~\inst{1}
    \and  P. Scheirich~\inst{1}
    \and  P.M. Garnavich~\inst{2}
    \and  S. Hameed~\inst{3}
    \and  D.A. Thilker~\inst{4} }


    \institute{Astronomical Institute, Academy of Sciences, CZ-251~65~Ond\v{r}ejov,
         Czech Republic, \email{k.hornoch@centrum.cz}
    \and University of Notre Dame, Department of Physics, 225 Nieuwland Science
         Hall, Notre Dame, IN 46556-5670, USA, \email{pgarnavi@nd.edu}
    \and Five College Astronomy Department, Smith College, Northampton,
         MA 01063, USA, \email{shameed@hampshire.edu}
    \and Center for Astrophysical Sciences, Johns Hopkins University,
         3400 North Charles Street, Baltimore, MD 21218, USA,
         \email{dthilker@skysrv.pha.jhu.edu}
    }
    \date{Received / Accepted 17 September 2008}

\abstract{}
{This paper reports on a search for new classical nova candidates in the M81 galaxy based on archival,
as well as recent, new images.}
{We used images from 1999--2007 to search for optical transients in M81. The positions of the
identified classical nova candidates were used to study their spatial distribution.
Kolmogorov - Smirnov test (KS) and bottom-to-top (BTR) ratio diagnostic were used to analyze
the nova candidate distribution and differentiate between the disk and the bulge populations.}
{In total, 49 classical nova candidates were discovered. In this study, we present the precise positions and
photometry of these objects, plus the photometry of an additional 9 classical nova candidates found
by Neill \& Shara (2004). With our large sample, we find a different spatial distribution
of classical nova candidates when compared to the results of earlier studies. Also, an extraordinarily 
bright nova was found and studied in detail.} {}

\maketitle

\section{Introduction}

Novae are important objects for the study of close binary
evolution, but our location in the Milky Way prevents us from
getting an unbiased sample locally. Studying novae in nearby galaxies can 
provide a more homogeneous sample of these
objects. For galaxies several Mpc away, long-term monitoring
coupled with a rapid cadence using a relatively large telescope is
necessary. Significant amounts of telescope time are difficult
to obtain  so searching for nova candidates using
archival images (already obtained for a variety of different science studies) have the
possibility for getting useful results. This method has several
disadvantages,including a lack of control over cadence, bandpasses, and exposure depth.

The outburst of classical novae (CNe) are caused by explosive hydrogen
burning on the white dwarf (WD) surface of a close binary system with material transfer from
the companion star onto the WD surface. During the thermonuclear runaway, a fraction
of the envelope is ejected, while a part of it remains in steady nuclear burning on the WD
surface (Jos\'e \& Hernanz 1998; Prialnik \& Kovetz 1995).
This powers a supersoft X-ray source (SSS). The duration of the SSS phase is inversely
related to the WD mass (Pietsch et al. 2006). Since WD envelope models also show
that the duration of the SSS phase depends on the metalicity of the envelope, monitoring
of the SSS phase of CNe also provides important information about the chemical composition
of the post-outburst envelope (Pietsch et al. 2006). Results of recent work aimed
at X-ray monitoring optical novae in M31 (Pietsch et al. 2006) bring new important
results, while showing the necessity of having a good catalog of optical novae
available for such studies.

Nearby galaxies with high annual nova rates are the best targets for statistically conclusive
studies of the properties of extragalactic novae at optical, as well as X-ray, wavelengths.
Besides M31, the M81 galaxy is another nearby large spiral galaxy. Only two recent papers aimed
at the study of CNe in M81 have been published up to now -- Shara et al. (1999)
and Neill \& Shara (2004). Here, we take advantage of M81 being a relatively common
target for optical imaging and, in order to search for classical nova (CN) candidates in this galaxy, we
analyze available archival CCD images, together with our recent images.
\section{Observations and data reduction}

\begin{table*}
\caption{Observers, observatory, telescopes and CCDs for measurements}
\label{tab1}
\centering                          
\scriptsize
\begin{tabular}{r c c c c}        
\hline \hline   
 ID & Observer & Observatory & Telescope & CCD \\   
\hline                        
(1) & S. Hameed, D. Thilker & KPNO & 4-m Mayall & Mosaic \\
(2) & P. Sorensen & La Palma & 2.54-m INT & WFC \\
(3) & M. Azarro & La Palma & 2.54-m INT & WFC \\
(4) & R. G. McMahon & La Palma & 2.54-m INT & WFC \\
(5) & S. Maddox & La Palma & 2.54-m INT & WFC \\
(6) & H. Deeg & La Palma & 2.54-m INT & WFC \\
(7) & A. Helmi & La Palma & 2.54-m INT & WFC \\
(8) & A. Herrero & La Palma & 2.54-m INT & WFC \\
(9) & L. C. Ho & & HST & ACS-WFC \\
(10) & S. J. Smartt & & HST & ACS-WFC \\
(11) & J. P. Huchra & & HST & ACS-WFC \\
(12) & Y. Taniguchi et al. & Mauna Kea & 8.3-m SUBARU & SuprimeCam \\
(13) & P. Garnavich et al. & Mt. Graham & 1.83-m VATT & VATT2K \\
(14) & J. E. Drew & La Palma & 2.54-m INT & WFC \\
(13) & P. Garnavich et al. & Mt. Graham & 1.83-m VATT & VATT2K \\
(14) & J. E. Drew & La Palma & 2.54-m INT & WFC \\
(15) & P. Garnavich, B. Tucker & KPNO & 3.5-m WIYN & Mini-Mosaic \\
(16) & Kaz, Yamada, Nakata & Mauna Kea & 8.3-m SUBARU & SuprimeCam \\
(17) & Arimoto, Ferguson, Jablonka & Mauna Kea & 8.3-m SUBARU & SuprimeCam \\
(18) & P. Garnavich & KPNO & 3.5-m WIYN & Mini-Mosaic \\
(19) & K. Nandra & La Palma & 4.2-m WHT & PFIP \\
(20) & J. Beckman & La Palma & 2.54-m INT & WFC \\
(21) & D. Christian & La Palma & 2.54-m INT & WFC \\
(22) & K. Hornoch & Ond\v{r}ejov & 0.65-m & AP7p \\
(23) & K. Hornoch & Lelekovice & 0.35-m & G2CCD-1600 \\
(24) & P. Caga\v{s} & Zl\'{\i}n & 0.26-m & G2CCD-3200 \\
(25) & P. Caga\v{s}, P. Caga\v{s}, Jr. & Zl\'{\i}n & 0.26-m & G2CCD-3200 \\
(26) & P. Caga\v{s}, V. P\v{r}ib\'{\i}k & Zl\'{\i}n & 0.26-m & G2CCD-3200 \\
(27) & V. P\v{r}ib\'{\i}k & Zl\'{\i}n & 0.26-m & G2CCD-3200 \\
(28) & L. Donato, G. Sostero & Remanzacco & 0.45-m & FLI-IMG 1001E \\
(28) & M. Gonano, V. Gonano & Remanzacco & 0.45-m & FLI-IMG 1001E \\
(29) & A. Lepardo, V. Santini & Remanzacco & 0.45-m & FLI-IMG 1001E \\
(30) & A. Tonelli & Roma & 0.13-m & SXV-H9 \\(31) & Inseok Song & Gemini Observatory & 8.1-m GEMINI North & GMOS-N \\
\hline                                  
\end{tabular}
\normalsize
\end{table*}

    Most of the images we used  were obtained from archives of large and medium-size
telescopes
such as Subaru, Gemini North, HST, the William Herschel Telescope and the Isaac Newton
Telescope. Archival
images were downloaded as raw FITS files and then processed, with the
exception of the images from
HST-ACS,
which were obtained as fully processed and calibrated .DRZ FITS files.
We also obtained recent images to add  substantially to our data
using medium-size
telescopes including Mayall 4-m, WIYN 3.5-m and VATT 1.83-m plus
additional images from small telescopes. Observatories, telescopes,
CCD cameras and names of observers for all the images used are given in Table~\ref{tab1}.

Using archival images originally obtained for different
purposes and taking data from many different telescopes brings a certain amount
of inhomogeneity into the data collected. The archival data have a variety of
fields of view (FOV), field centers, limiting
magnitudes and passbands which complicate the analysis. Most of the images used come
from the 2.54-m Isaac Newton Telescope at the La Palma and these images were
taken using a wide-field CCD camera with the FOV
about half a degree wide, centered mostly on the core of the M81
galaxy. They, as well as our recent images from the
Mayall 4-m, cover whole galaxy. Our images from the WIYN 3.5-m with a 
smaller FOV of 9.7\arcm $\times$ 9.7\arcm\ cover the central part of the galaxy.

The images from the Subaru, Gemini, WHT and VATT
are not centered on the galaxy center but in all cases they cover considerable
parts of the M81 galaxy, including the nucleus. The archival images from the HST-ACS
taken in the wide-field mode are a special case. They cover relatively small area
of 3.4\arcm $\times$ 3.4\arcm\ of the central part of the galaxy. Generally, the spatial
coverage of the entire galaxy by the images used is rather uniform, with an exception
of outer parts of the galaxy, where coverage by images is less frequent when compared with
the central region.

    Most of the images were taken using the narrow-band $H_{\alpha}$ and broad-band
R (or SDSS r') filters. Also, B and V filtered images were
used, as well as HST-ACS F814W and F658N filtered images. In the special case
of the two CN candidates M81N~2007-04a and M81N~2007-04b, unfiltered images were also
used, having been taken using small telescopes; unfiltered images were necessary to
achieve a sufficient signal-to-noise ratio (S/N).

    Standard reduction procedures for raw CCD images were applied (bias and
dark-frame subtract and flat-field correction) using SIMS\footnote{\tt http://ccd.mii.cz/}
and Munipack\footnote{\tt http://munipack.astronomy.cz/} programs.
Reduced images of the same series were co-added to improve the S/N ratio
(total exposure time varied from a few minutes up to about three hours).
The gradient of the galaxy background of co-added images was flattened by the spatial
median filter using SIMS. These processed images were used to search for
nova candidates, photometry and astrometry.

\begin{figure*}[]
\clearpage
\includegraphics[angle=0, width=18cm, clip]{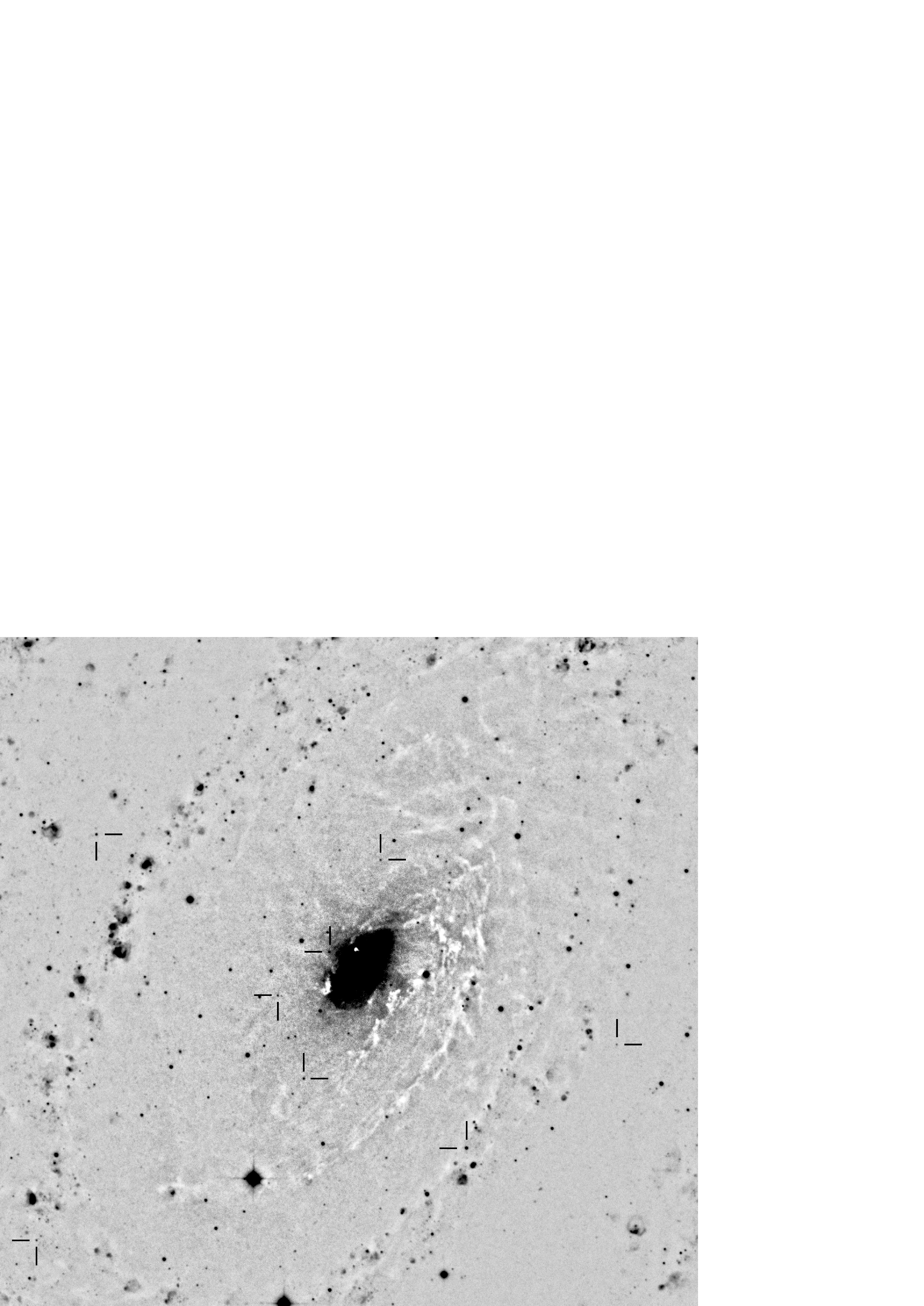}
\centering
\caption[]{Eight nova candidates in M81: Example of high-quality narrow-band
$H_{\alpha}$ image used in our search for novae. Original images were obtained
under excellent seeing by P. Sorensen using the 2.5-m Isaac Newton Telescope at
La Palma. Raw images downloaded from the Isaac Newton Group Archive were processed
using standard reduction procedures and co-added; spatial median filter was then applied
on co-added image. Image presented here has 1800 seconds of total exposure time
and shows 7\arcm $\times$ 7\arcm of the central region of M81. Each of eight
classical nova candidates recorded on this image are marked by two perpendicular
lines.}
\label{9592fig1}
\end{figure*}

\subsection{Searching for novae}

 The search for nova candidates was performed by means of visual comparison of
an image with the best available image taken in
the same passband with a substantial separation in time.
A majority of the archival images were taken in the narrow-band
$H_{\alpha}$ filter, so we were able to construct
a deep $H_{\alpha}$ image by co-adding many 2.54-m Isaac Newton Telescope images
and this image was
then used as a ``master'' comparison for searching through other $H_{\alpha}$
images. To confirm their transient nature, the nova candidates were required to be missing
on images with sufficiently deep limiting magnitude, taken in epochs out of the span
of the observed outburst.

    When a transient was found, we inspected all single images used for
co-added frames to exclude the possibility of cosmic ray hit or any defect on the
CCD detector or processing artifact. Brightness limits were estimated on
deep images for all nova candidates to confirm their transient nature. 
Once the transient's existence is confirmed, we performed photometry and astrometry.

For a transient to be  classified as a CN candidate its absolute magnitude
had to be sufficiently high, and its time span
of observability had to be relatively short ($\ll 1$ year) in available images of the period 1999--2007.
We found variable objects that do not meet these thresholds and such objects
are not presented here.

\subsection{Photometry}

When an object was classified as CN candidate, photometry
and astrometry were done following as follows. ``Optimal photometry'' (based on fitting of PSF
profiles) using GAIA\footnote{\tt http://www.starlink.rl.ac.uk/gaia} and 
astrometry using APHOT (a synthetic aperture photometry and
astrometry software developed by M. Velen and P. Pravec at the Ond\v{r}ejov observatory, see Pravec et al. 1994)
were performed.
   
 For narrow-band $H_{\alpha}$ photometry we calibrated 17 stars in the M81
field using the white dwarf HZ~44 $\alpha_\mathrm{J2000} = 13^h23^m35\fs37,
\delta_\mathrm{J2000} = +36\degr08\arcmin00\farcs0$
as a spectrophotometric standard star (Massey \& Strobel~1988, Landolt \& Uomoto~2007).
$B,V$ and $R$
magnitudes for comparison stars located in the M81 field were taken from Perelmuter \& Racine (1995).
The HST-ACS images were an exception - the magnitudes in the STMAG photometric system were derived
using the standard procedure for HST-ACS drizzled (DRZ) images described by Sirianni et al. (2005).

\subsection{Astrometry}

    For astrometry, we created (with help of M.~Velen and P.~Pravec) a
special catalog of fainter stars in the M81 field derived from the
2.5-m INT telescope images corrected to the world coordinate system (WCS)
using stars from the USNO-A 2.0 catalog. Depending on the field of view, 
tens to more than 300 stars from this catalog were used for
WCS mapping of images.
The mean residual of catalog positions is about 0.2\arcs and the nova
positions derived have uncertainties in order of tenths of arcsecond.

\section{Results}

In total, we classified 49 transient objects as CN candidates.
We also independently detected
additional 9 novae from the total number of 12 novae already found by Neill \& Shara
(2004). The photometry results for our 49 CN candidates are given in Table~\ref{tab4},
while the results for the 9 novae already found by Neill \& Shara (2004) are summarized
in Table~\ref{tab5}. Typical errors of photometry are 0.1~--~0.2 mag.
The precise positions, offsets
from the M81 center, designations and discoverers of 49 CN
candidates are given in Table~\ref{tab3}.

\begin{table*}
\begin{center}
\caption[]{Nova candidate designations, positions and discoverers.
 For offset we used $\alpha_\mathrm{J2000} = 9^h55^m33\fs173,
 \delta_\mathrm{J2000} = +69\degr03\arcmin55\farcs06$ as reference position of
the M81 center. Discoverers for individual nova candidates are given under the column ``Disc.''.
For photometric data of individual nova candidates, see Table~\ref{tab4} }
\label{tab3}
\scriptsize
\begin{tabular}{llccrrc}
\hline\noalign{\smallskip}
\hline\noalign{\smallskip}
No. & Name$^a$ & R.A.  & Decl. & Offset & &  Disc.$^b$ \\
    & M81N     & J2000 & J2000 & R.A.   & Decl. &   \\
\noalign{\smallskip}\hline\noalign{\smallskip}
1999-1 & 1999-11a & 09 55 34.11 & +69 01 49.1 &   5.0\arcs E & 126.0\arcs S  & (c) \\
1999-2 & 1999-11b & 09 54 56.16 & +69 03 56.8 & 198.4\arcs W &   1.7\arcs N  & (c) \\
2000-1 & 2000-11a & 09 55 29.50 & +69 02 52.4 &  19.7\arcs W &  62.7\arcs S  & (c) \\
2000-2 & 2000-12a & 09 55 54.31 & +69 00 50.7 & 113.4\arcs E & 184.4\arcs S  & (a) \\
2001-1 & 2001-01a & 09 55 21.08 & +69 02 06.4 &  64.9\arcs W & 108.7\arcs S  & (b) \\
2001-2 & 2001-01b & 09 56 03.25 & +69 05 12.2 & 161.1\arcs E &  77.1\arcs N  & (b) \\
2001-3 & 2001-01c & 09 55 42.54 & +69 03 36.8 &  50.2\arcs E &  18.3\arcs S  & (b) \\
2001-4 & 2001-01d & 09 55 39.49 & +69 02 46.8 &  33.9\arcs E &  68.3\arcs S  & (b) \\
2001-5 & 2001-01e & 09 55 36.77 & +69 04 03.5 &  19.3\arcs E &   8.4\arcs N  & (b) \\
2001-6 & 2001-01f & 09 55 31.12 & +69 04 59.5 &  11.0\arcs W &  64.4\arcs N  & (b) \\
2001-7 & 2001-01g & 09 56 09.15 & +69 01 06.4 & 193.0\arcs E & 168.7\arcs S  & (b) \\
2001-8 & 2001-01h & 09 55 04.31 & +69 03 10.0 & 154.7\arcs W &  45.1\arcs S  & (b) \\
2002-1 & 2002-12a & 09 56 03.50 & +69 02 50.4 & 162.6\arcs E &  64.7\arcs S  & (a) \\
2002-2 & 2002-12b & 09 55 32.30 & +69 03 50.0 &   4.7\arcs W &   5.1\arcs S  & (a) \\
2002-3 & 2002-12c & 09 55 28.30 & +69 03 39.9 &  26.1\arcs W &  15.2\arcs S  & (a) \\
2002-4 & 2002-12d & 09 55 20.81 & +69 02 56.5 &  66.3\arcs W &  58.6\arcs S  & (a) \\
2002-5 & 2002-12e & 09 55 15.68 & +69 03 23.5 &  93.8\arcs W &  31.6\arcs S  & (a) \\
2003-1 & 2003-05c & 09 56 07.47 & +69 03 57.0 & 183.8\arcs E &   1.9\arcs N  & (a) \\
2003-2 & 2003-09a & 09 55 40.76 & +69 02 35.5 &  40.7\arcs E &  79.6\arcs S  & (a) \\
2003-3 & 2003-09b & 09 55 53.81 & +69 02 10.6 & 110.7\arcs E & 104.5\arcs S  & (a) \\
2003-4 & 2003-09c & 09 55 33.27 & +69 03 37.5 &   0.5\arcs E &  17.6\arcs S  & (a) \\
2003-5 & 2003-05a & 09 55 27.19 & +69 04 17.1 &  32.1\arcs W &  22.0\arcs N  & (a) \\
2003-6 & 2003-05b & 09 54 53.85 & +69 01 11.1 & 211.0\arcs W & 164.0\arcs S  & (a) \\
2004-1 & 2004-02c & 09 55 34.71 & +69 04 20.3 &   8.2\arcs E &  25.2\arcs N  & (a) \\
2004-2 & 2004-02a & 09 55 27.47 & +69 04 51.6 &  30.6\arcs W &  56.5\arcs N  & (a) \\
2004-3 & 2004-02b & 09 55 33.82 & +68 58 32.6 &   3.5\arcs E & 322.5\arcs S  & (a) \\
2004-4 & 2004-09a & 09 55 43.22 & +69 03 40.9 &  53.9\arcs E &  14.2\arcs S  & (a) \\
2005-1 & 2005-12a & 09 56 15.23 & +69 00 56.9 & 225.7\arcs E & 178.2\arcs S  & (a) \\
2005-2 & 2005-12b & 09 55 33.74 & +69 05 40.5 &   3.0\arcs E & 105.4\arcs N  & (a) \\
2005-3 & 2005-01a & 09 56 08.74 & +69 04 24.3 & 190.6\arcs E &  29.2\arcs N  & (a) \\
2005-4 & 2005-11a & 09 55 27.53 & +69 04 57.3 &  30.2\arcs W &  62.2\arcs N  & (a) \\
2006-1 & 2006-02a & 09 55 35.49 & +69 04 08.6 &  12.4\arcs E &  13.5\arcs N  & (b) \\
2006-2 & 2006-02b & 09 55 33.98 & +69 03 36.3 &   4.3\arcs E &  18.8\arcs S  & (b) \\
2006-3 & 2006-02c & 09 55 15.16 & +69 06 12.1 &  96.5\arcs W & 137.0\arcs N  & (b) \\
2006-4 & 2006-02d & 09 55 40.76 & +69 03 31.1 &  40.7\arcs E &  24.0\arcs S  & (b) \\
2006-5 & 2006-02f & 09 55 30.40 & +69 03 35.6 &  14.9\arcs W &  19.5\arcs S  & (b) \\
2006-6 & 2006-02e & 09 55 12.21 & +69 00 19.1 & 112.5\arcs W & 216.0\arcs S  & (b) \\
2006-7 & 2006-12a & 09 55 41.29 & +69 04 20.8 &  43.5\arcs E &  25.7\arcs N  & (d) \\
2006-8 & 2006-12b & 09 55 43.01 & +69 03 56.8 &  52.7\arcs E &   1.7\arcs N  & (d) \\
2006-9 & 2006-03a & 09 55 50.82 & +69 04 03.8 &  94.6\arcs E &   8.7\arcs N  & (a) \\
2006-10& 2006-03b & 09 55 39.52 & +69 01 33.3 &  34.0\arcs E & 141.8\arcs S  & (a) \\
2006-11& 2006-02g & 09 55 28.62 & +69 01 51.2 &  24.4\arcs W & 123.9\arcs S  & (b) \\
2006-12& 2006-01a & 09 55 44.16 & +69 02 58.4 &  58.9\arcs E &  56.7\arcs S  & (a) \\
2006-13& 2006-02h & 09 55 35.84 & +69 03 55.7 &  14.3\arcs E &   0.6\arcs N  & (a) \\
2006-14& 2006-02i & 09 55 36.27 & +69 03 11.2 &  16.6\arcs E &  43.9\arcs S  & (a) \\
2006-15& 2006-02j & 09 55 31.59 & +69 02 51.7 &   8.5\arcs W &  63.4\arcs S  & (a) \\
2007-1 & 2007-01a & 09 55 19.75 & +69 03 26.0 &  72.0\arcs W &  29.1\arcs S  & (c) \\
2007-2 & 2007-04a & 09 55 28.58 & +69 04 21.6 &  24.6\arcs W &  26.5\arcs N  & (e) \\
2007-3 & 2007-04b & 09 55 31.30 & +69 05 28.9 &  10.0\arcs W &  93.8\arcs N  & (f) \\
\hline
\end{tabular}

\medskip
Notes: \hspace{0.05cm} $^a$: following CBAT nomenclature for novae in M31
(see {\tt http://cfa-www.harvard.edu/iau/CBAT\_M31.html}) \\
\hspace*{0.95cm} $^b$: Discoverers of novae: (a) K. Hornoch, (b) K. Hornoch, S. Hameed, D. Thilker, (c) K. Hornoch \& P. Garnavich, \\
\hspace*{1.25cm} (d) K. Hornoch, P. Garnavich, B. Tucker, (e) K. Hornoch, P. Caga\v{s}, P. Caga\v{s}, Jr., (f) V. P\v{r}ib\'ik, K. Hornoch, P. Caga\v{s} \\
\end{center}\end{table*}

Finder charts for all of 49 objects reported in this paper are accessible at the
Supernovae webpage\footnote{\tt http://www.supernovae.net/novae.html}.
Positions of these objects, together with positions of CN candidates from other studies are shown 
in Fig.~\ref{9592fig7}.
We note that the major axis of the galaxy is plotted by solid line. We adopt
$150\degr$ for the position angle of the major axis of M81 (see Goad 1976).

\begin{figure}[!h]
\includegraphics[angle=270, width=0.48\textwidth]{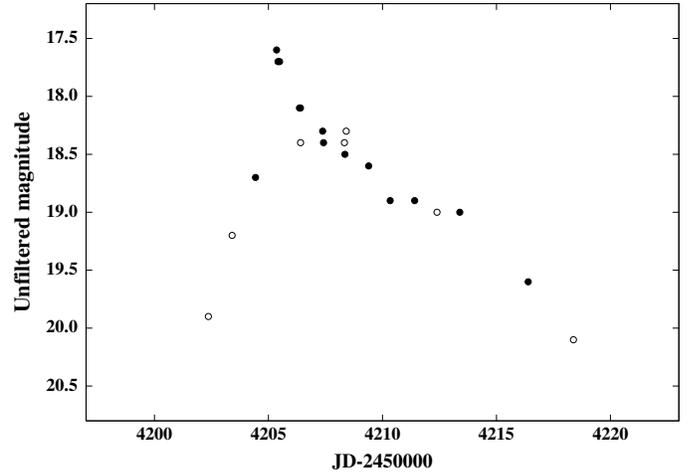}
\caption[]{
Light curve of an exceptionally bright nova M81N~2007-04b. Filled and open circles indicate measurements
with an uncertainty $\leq$ 0.2 mag and $>$ 0.2 mag, respectively.}
\label{9592fig2}
\end{figure}

\begin{figure}[!h]
\includegraphics[angle=270, width=0.48\textwidth]{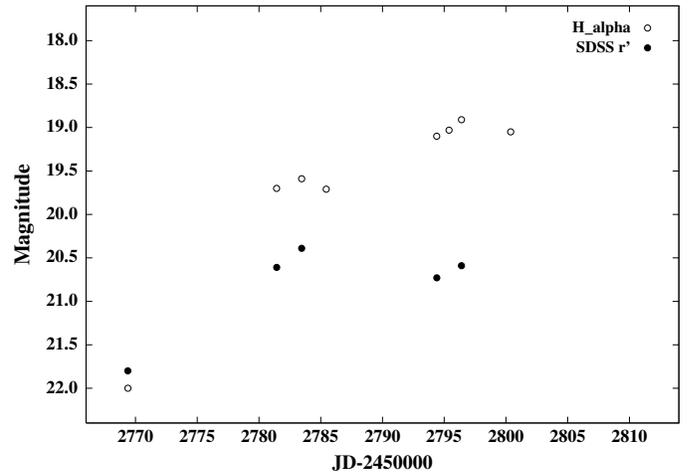}
\caption[]{
Light curve of classical nova candidate M81N~2003-05c: A typical light curve for a slow nova
with long-lasting raise phase $>$ 25 days in $H_{\alpha}$ showing its raise phase. }
\label{9592fig3}
\end{figure}

In the course of our survey, we have discovered a very interesting nova M81N~2007-04b
in images taken by P.~Caga\v{s} with the 0.26-m telescope at the Zl\'{\i}n observatory.
Immediately after the discovery, we started a campaign for monitoring this exceptionally
bright nova. We carried out unfiltered photometry using relatively small telescopes.
Co-added images with typical exposure times of about 1.5~--~2 hours enabled us to carry
out photometry with a relative high precision (typical errors 0.1~--~0.15 mag). The
light curve has very good temporal coverage and shows a relatively slow rise
and then a typical decline phase from the maximum light for a fast nova (see Fig.~\ref{9592fig2}).
Using our photometry we derived the rate
of decay of 0.17 magnitudes per day in good agreement with ~0.15 magnitudes per day derived
from the decline phase obtained from the photometry in SDSS r' filter and published
by Rau et al. (2007). The time interval of decay by 2 magnitudes from the maximum
light $T_{2mag} = 12$ days; it suggests that this object falls into the group of fast novae.
Three spectra of this nova were obtained using the 10-m Keck I telescope, one at the
maximum light phase by Silverman et al. (2007) and two by Rau et al. (2007) five days
later, suggesting type FeIIn spectral classification. This nova became one of the brightest
and most thoroughly studied novae in the M81 galaxy and it is probably only the second nova
in M81 studied spectroscopically.
As an example of a slow nova can serve M81N~2003-05c; SDSS r' and narrow-band $H_{\alpha}$
photometry were obtained, as shown in Fig.~\ref{9592fig3}. Although it shows an incomplete
light curve, very slow brightness changes are clearly visible -- the rising phase in
narrow-band $H_{\alpha}$ lasted $>$ 25 days. Also it is clearly visible that the maximum
phase in the SDSS r' band (which includes entire $H_{\alpha}$ emission) was reached
significantly earlier than in the narrow-band $H_{\alpha}$ which is typical for classical
nova explosions, in which maximum intensity of $H_{\alpha}$ emission occurs after the maximum
light phase in continuum.

\subsection{Spatial distribution}

	We used two simple methods to describe a spatial distribution of our CN candidates.
In the first method we compared cumulative radial distributions of the bulge, the disk, and
the total light of M81 with the cumulative radial distribution of our CN candidates.
The second method is based on a bottom-to-top ratio of novae in the galaxy (see below).
All 49 CN candidates from this work plus 9 additional novae independently found on our images
and already published by Neill \& Shara (2004) were used.

\begin{figure}[!h]
\includegraphics[width=0.48\textwidth]{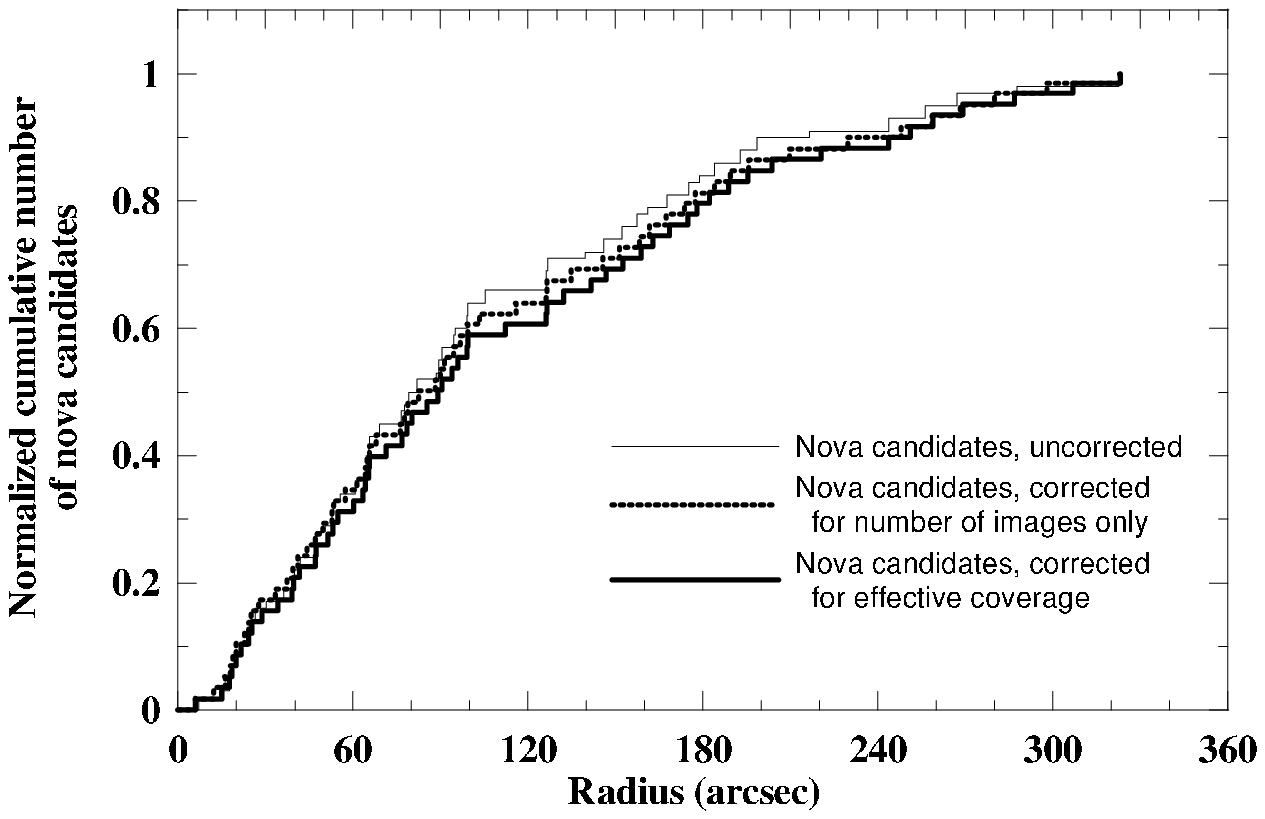}
\caption{Cumulative radial number distribution of nova candidates. See text for details.}
\label{9592fig4}
\end{figure}

\begin{figure}[!h]
\includegraphics[width=0.48\textwidth]{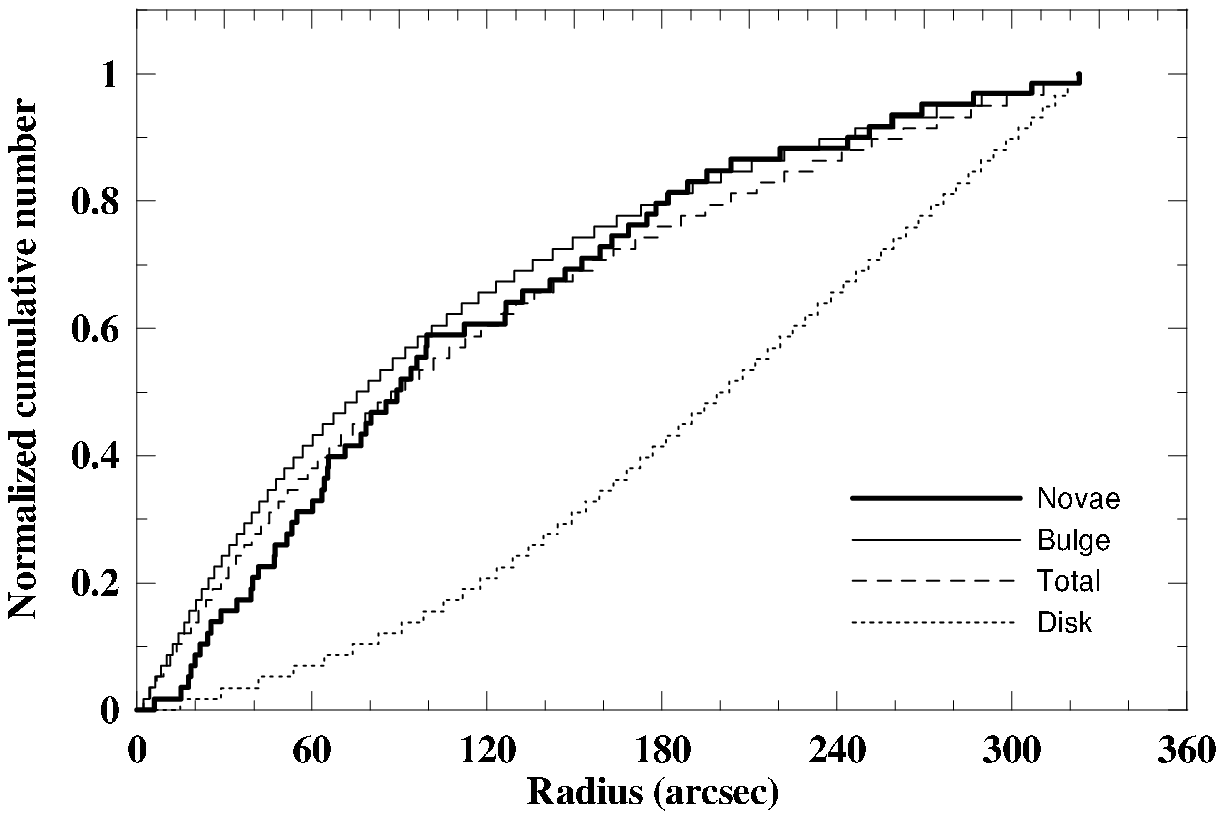}
\caption{The cumulative radial distribution of nova candidates, corrected for effective
coverage, compared with the components of galaxy light. See text for details.}
\label{9592fig5}
\end{figure}

In the first method, we used the same cumulative radial distribution for the bulge, the disk, and the total
light, as Neill \& Shara (2004) by fitting the curves in their Fig.~10 to spline functions.
These cumulative radial distributions, together with the cumulative radial distribution of our CN
candidates, corrected for effective coverage (see below and Fig.~\ref{9592fig4})
are plotted in Fig.~\ref{9592fig5}.

We did not detected any CN candidates beyond about 5\arcm from the galaxy center, so
we normalized all distributions to unity at radius of 323\arcs, where the farthest CN candidate was detected.
A most probable explanation for why we did not detect any CN candidates in the outer regions of the galaxy
is a sparse time coverage of the images covering the whole galaxy (see Section~\ref{gaps}).
Corrected and raw cumulative radial distributions of CN candidates were constructed
and are plotted
in Fig.~\ref{9592fig4} and show that differences among them are small. So we conclude that
selection effects (up to radius of 323\arcs) do not influence the results significantly.
The raw cumulative radial distribution of the CN candidates without correction is shown
as a thin solid line.
A second radial distribution was constructed
by counting the contribution of each nova candidate as $1/N$, where $N$ is the number of images
that include the nova's position. In a third distribution, the radial distribution
of candidates was corrected by calculating the effective coverage $N_{eff}$ (see Section.~\ref{EffCover}).
The $1/N$ corrected and the effective coverage corrected cumulative radial distributions
are plotted in Fig.~\ref{9592fig4} as bold dotted and bold solid lines, respectively.

We use the Kolmogorov--Smirnov (KS) test to quantify the similarity between the cumulative radial
distribution of CN candidates and the cumulative radial distributions of three galaxy light distributions.
The best match with the distribution of CN candidates is the total light distribution which can be ruled out only
at the 2.3\% confidence level. The bulge and the disk light distributions can be ruled out at the 24\% and $99.99\%$
confidence levels, respectively. The excellent match between the total light distribution indicates that both
the disk and the bulge are contributing novae to our sample.

As the second method for diagnostic of the bulge-to-disk ratio of novae we used a method described
in Hatano et al. (1997) and used in the previous works by Neill \& Shara (2004) and Shara et al. (1999),
which, however, differ in their results. The method evaluates the effect of dust
on the distribution of observed novae in the bulge region. If novae are present
primarily in the bulge, a large asymmetry in their distribution would be visible because the dust in
the disk obscures apparent bulge novae behind the disk. On the other hand, if novae arise primarily
in the disk, only small asymmetry should be present. The major axis of the galaxy is used as a dividing
line, and the numbers of novae above and below this line are compared.
The ratio is called a bottom-to-top ratio (BTR).

For the full sample of 58 CN candidates, we divide the galaxy along the major axis into the north-east
side (top of the bulge) and into the south-west side (bottom of the bulge). We also separate these
subsamples into CN candidates
within 90\arcsec, 150\arcsec, and 210\arcsec from the nucleus of M81. These numbers
were then corrected for effective coverage and are presented in Table~\ref{tab6} together with
uncorrected numbers and corresponding BTRs. The numbers of novae and BTRs from Shara et al. (1999),
and Neill \& Shara (2004) were counted for the same distances from the centre, and are presented
in the table as well.

The BTR from our data is close to unity and is roughly independent of radial distance.
This fact is important in this method, and probably it is caused by a higher number of CN
candidates in our dataset in comparison with previous works.
Also, correction for effective coverage caused  insignificant changes to the BTR, and thus,
inhomogeneity in the spatial coverage present in our dataset did not strongly affect the results.

The fact that the asymmetry in spatial distribution of our CN candidates is low (corresponding to BTR close to~1)
supports the hypothesis of significant contribution of the disk population of novae.

In previous work, the BTR varies from 1.0 to 1.8 and from 0.25 to 0.5 (Shara et al. 1999,
Neill \& Shara 2004, respectively). The results from the former indicated low bulge-to-disk nova
ratio, whereas Neill \& Shara (2004) concluded that the bulge-to-disk nova ratio is very high.
However, these studies suffer from incompleteness in the bulge
area (Shara et al. 1999), and a small sample size (Neill \& Shara 2004), and these problems could have
resulted in the inconsistent results.

It should be noted that while our results are based on an assumption that the dust
distribution in M81 is similar to that in M31, their interpretation has to be done with caution.
The fact that M81 is slightly more face-on than M31 should slightly reduce the asymmetry
in the bulge-to-disk nova ratio (e.g., a face-on galaxy would show no asymmetry regardless of the
bulge-to-disk ratio at all). A detailed model of dust distribution in M81 would be necessary for making
more quantitative conclusions from our results.

\begin{table*}
\begin{center}
\caption[]{Numbers of CN candidates located bottom of the bulge, top of the bulge, and bottom-to-top
ratio (BTR), within 90\arcsec, 150\arcsec, and 210\arcsec, respectively, of the nucleus of M81
from this work, Shara et al. (1999), and Neill \& Shara (2004). See text for details.}
\label{tab6}
\scriptsize
\begin{tabular}{lccccccccccc}
\hline\noalign{\smallskip}
\hline\noalign{\smallskip}
 & & d $\leq 90\arcsec$ & & & & d $\leq 150\arcsec$ & & & & d $\leq 210\arcsec$ \\
 & & & & & & & & & & & \\
 & Bottom & Top & BTR & & Bottom & Top & BTR & & Bottom & Top & BTR \\
\noalign{\smallskip}\hline\noalign{\smallskip}
This work, uncorrected & 16.0 & 16.0 & 1.00 & & 21.0 & 22.0 & 0.95 & & 25.0 & 27.0 & 0.93 \\
This work, corrected for effective coverage & 16.2 & 16.2 & 1.00 & & 21.3 & 22.3 & 0.96 & & 25.3 & 27.4 & 0.92 \\
Shara et al. (1999) & 1.0 & 1.0 & 1.00 & & 4.0 & 4.0 & 1.00 & & 9.0 & 5.0 & 1.80 \\
Neill \& Shara (2004) & 2.0 & 4.0 & 0.50 & & 2.0 & 8.0 & 0.25 & & 3.0 & 8.0 & 0.38 \\

\hline\noalign{\smallskip}
\end{tabular}
\end{center}\end{table*}

\subsection{Correction for effective coverage} \label{EffCover}

A probability of nova discovery obviously depends on a number of images covering the nova
position. In order to eliminate these selection effects, we created a map in R.A. and declination
offset describing a coverage of the galaxy and its surroundings. For each point of this map,
the number of images, $N = \sum_i Q_i$, covering this point was computed, where $Q_i = 1$ if the $i$-th
image includes the point and $0$ if it doesn't. Since the limiting magnitude ($LM$) varies for
different images, when counting the number of images, contribution of each image was weighted
by a factor $f(LM)$, so that the total effective coverage of the points was computed as
\begin{equation}
N_{eff} = \sum_i (Q_i S_i)/f(LM_i),\label{Neff} \label{NEff}
\end{equation}
where $LM_i$ is the limiting magnitude of $i$-th image.
The factor $S_i$ in the above formula was set to $0$ or $1$, depending on whether the $i$-th
image was saturated at the position of examined point or not.

The $LM$ symbol in this section refers to limiting magnitude in narrow-band $H_{\alpha}$ filter.
Due to a low cadency of images in our survey, we suppose that most of the novae was
detected after they reached maximum of brightness; we used the $H_{\alpha}$ passband for $LM$ since the novae have
strong emission in $H_{\alpha}$ during their decline phase.
Since about half of the images was not taken in $H_{\alpha}$ filter,
the limiting magnitudes estimated for these images were transformed to $LM$ in $H_{\alpha}$ by subtracting
of 2.5 mag for unfiltered images and images taken in B and V, and of 2 mag for images taken in R and SDSS r' filters.

The function $f(LM)$ is defined as dependence of a number of images required to detect one CN candidate
on a limiting magnitude $LM$.
The function was estimated empirically by fitting four pairs of values $[f_j, LM^{aver}_j]$, $j=1 \dots 4$,
with
\begin{equation}
f(LM) = 4.991 \times 10^{11} (LM)^{-8.9955},
\end{equation}
where $LM^{aver}_j$, $j=1 \dots 4$, are values averaged from $LM$ of images obtained from INT, Gemini, WIYN
and Ond\v{r}ejov 0.65-m telescopes.
The values $f_j$ are computed as $n_j/k_j$, where $n_j$ and $k_j$ are numbers of images taken and numbers of CN candidates
discovered by $j$-th telescope, respectively.
For most of images from each of four telescopes separately, differences between $LM$ of individual images and $LM^{aver}$
are lower than 1 mag.
In case of INT, Gemini and WIYN we used images taken in narrow-band $H_{\alpha}$ filter,
the images from 0.65-m telescope were unfiltered, but their limiting magnitudes were corrected to $LM$ as is described above.

\begin{figure}[!h]
\includegraphics[width=0.48\textwidth]{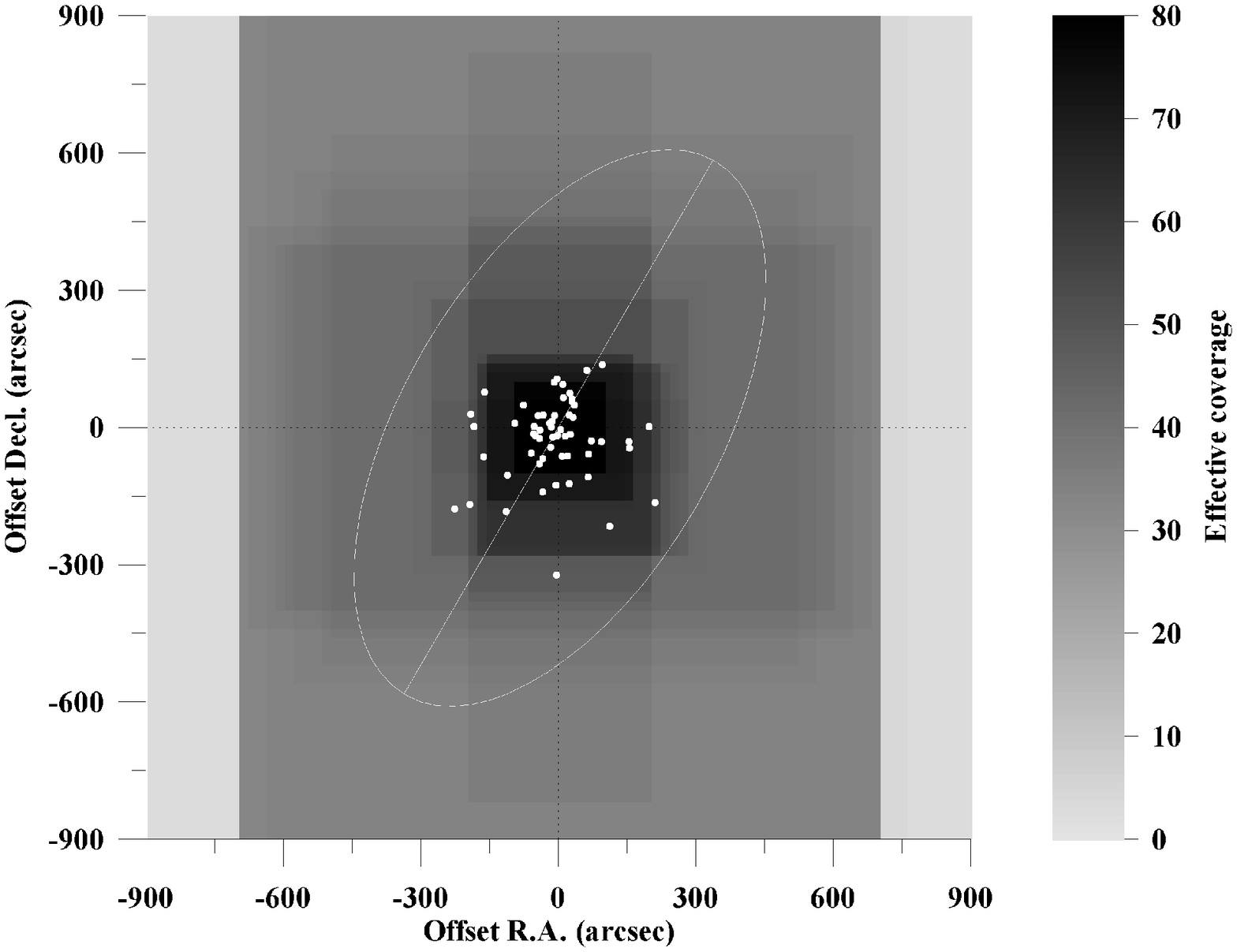}
\caption{Effective coverage of the images used in this work. Outer boundary of the M81
galaxy and its major axis are plotted as solid line. The classical nova candidates from
this work plus 9 novae found by Neill \& Shara (2004) are plotted as white circles.
See text for details.}
\label{9592fig6}
\end{figure}

\begin{figure*}[!t]
\includegraphics [angle=270, width=12cm]{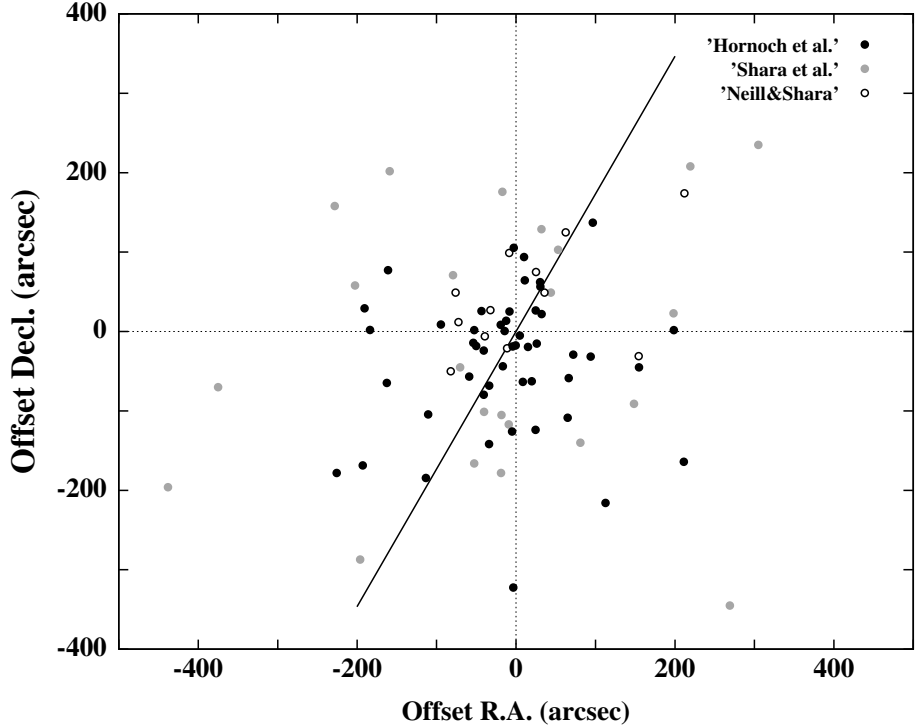}
\centering
\caption[]{Spatial distribution of 84 classical nova candidates in M81:
 Classical nova candidates found by Shara et al. (1999), by Neill \& Shara
 (2004) and from this work.}
\label{9592fig7}
\end{figure*}

Resulting map of the effective coverage $N_{eff}$ is shown on Fig.~\ref{9592fig6}.
We note that maximum value of $N_{eff}$ over whole map differs by 27\% from maximum value of $N$,
and a ratio between maximal effective coverage and the effective coverage in the outer regions of the galaxy is 2.3.

\section{Discussion}
While we present a large number of CN candidates, due to our limited temporal coverage,
some of our nova candidates are detected on images from only one night. Although positive
detection in at least two nights are commonly required, we also present here objects which
do not meet this requirement.
Here we consider possible sources of bias in cases when only one positive detection of the nova
candidate was made. Two types of bias were taken into account -- the first are image artifacts,
the second are real objects. Image artifacts with star-like appearance can be caused by cosmic
ray hits (radiation events) and by defects on the CCD chip. From the class of real objects,
we took into account Kuiper belt objects and variable stars.
    Cosmic ray hits usually have (mainly on well-sampled images) a different PSF profile and
as such are reliably identified only on one image from a series of images used for a co-added
image. This enables us to remove such cases relatively simply.

    CCD chip artifacts (defects) may produce false objects, even if correct image processing
procedures are used. Checking every image from the series may not identify such cases reliably,
because many of the images we used  were guided and with no shifts between them, so we also
checked the PSF profile of suspicious objects in the co-added image.

\subsection{Kuiper belt objects}

    Kuiper belt objects (KBOs) are particular class of real objects which can be potentially
considered as CN candidates when an image from one night only is available. A relatively small
apparent motion (up to about 0.06\arcs/min) can produce star-like appearance in an image exposed
during an interval in the order of minutes, even if a large telescope is used. Although the chance
of detecting a sufficiently bright KBO in a relatively small field, furthermore placed at a high
ecliptic latitude, is pretty small, we can estimate an upper limit of probability.
All objects detected in only one night were brighter than $R=22$, mag with an exception
of the object M81N~2004-09a. This object was found on the HST-ACS F814W filtered image and thanks
to the very good resolution of this image, an apparent motion of a typical KBO should be visible.
Moreover, the probability that some KBO brighter than $V=23.5$ mag at this position will fall into
small FOV of HST-ACS is very low. We assume that the total number of KBOs brighter than $V=22.5$ mag
(which corresponds to $H=5.5$ at the distance of 50 AU) is $< 100$ (based on Jedicke et al., 2002).
A typical image covers approximately 400\arcm$^2$ and we used images from 60 epochs
in total, so the probability that we detect a KBO brighter than $V = 22.5$ mag is $< 0.013$, which
is the hard upper limit. If we take into account that many of them would be possible to distinguish
using other images taken in similar epochs and that KBOs are concentrated to small ecliptic latitudes
(ecliptic latitude of M81 is 51.6 deg), it is very unlikely that some of our CN candidates
can really be KBOs.

\subsection{Variable stars}

    Estimating possible contamination of our sample of M81 CN candidates by other types of variable
stars presents a difficult problem. Objects detected on single epochs, as well as those detected in
multiple nights could be confused with non-novae. All objects that were classified as CN candidates
have $R = 21.5$ mag or brighter on available images with the exception of objects M81N~2004-09a and
M81N~2005-01a which have recorded magnitudes brighter than  $R = 23$ and 22.5, respectively.
Assuming a distance modulus of 27.8 mag for M81, the magnitude limits correspond to $M_R \sim -6$ mag
(and $-4.5$ to $-5$ mag) for most candidates (M81N~2004-09a and M81N~2005-01a, respectively).
We may discard types of variable stars which remain fainter than $M_R = -6$ mag. Thus, only Luminous
Blue Variables (LBV), Yellow Supergiants (YHG), Cepheids and supernovae remain as eligible contaminants.

    Considering their typical brightness, LBV stars should be visible most of the time above the limit
of most CCD frames used. This behavior does not meet with our CN candidates which were visible only
during a limited period.

    YHG stars have very small amplitudes which excludes this type of variables as a possible source
of CN candidates contamination.

    Cepheids have a sufficiently high absolute magnitude to be recorded in our images, but they have
relatively small amplitude (around 2 magnitudes), which makes the possibility of confusing some of the Cepheids
with our CN candidates relatively small. Moreover, Cepheids should be recorded several times (since they
are reaching at least -4.5 mag of absolute magnitude) over more than 7-year period covered by images.

Although supernovae (SNe) have more than sufficient luminosity to be recorded in our data, the
possibility of contamination by M81 SNe is extremely improbable for a number of reasons, such as
very low production of SNe in any galaxy (one per tens of years at best) and the bright apparent magnitude
around maximum light phase, thus making a SN in M81 very easy to detect around its maximum well before
their brightness decreases into the range of our CN candidates. Moreover, SNe have a very slow brightness
decay after fading many magnitudes below the maximum -- this makes it impossible to confuse such object
in our group of CN candidates.

Distant background SNe are more likely to invade the CN sample than supernovae in M81.
We estimated the number of background SNe which could contaminate our sample of CN candidates using
results from the Sloan Digital Sky Survey (SDSS). They found approximately 500 SNe brighter than
$r' = 22$ mag (it is similar to a typical limiting magnitude of our images) over 9 months of searching
300 square degrees (Frieman et al. 2008), i.e. 0.2 SNe per square degree per month. If our typical M81
image covers 0.11\degr$^2$, we are expecting only $\sim 0.02$ SNe per month brighter than $r' = 22$ mag
behind M81. Thus, we estimate that we could have recorded about two SNe in the interval of 90 months
covered by our images. Furthermore, one can expect that light from background SNe to be obscured by dust
in M81 so the number of background SNe brighter than our typical limiting magnitude should be even a bit
lower than we estimated above (gaps in time coverage by our images in order of months could only decrease
a probability of detection of background SN too). In the course of searching for CN candidates one probable background SN located close to an outer edge of M81 (exploded in faint anonymous galaxy with magnitude
of $R = 20.1$) was found by Hornoch (2008). Based on these results, we conclude that the contamination
of our sample of CN candidates by background SNe is very low, with a probable number of zero.

Another possible contamination of our sample of CN candidates could be foreground (Galactic) variable stars
that could be confused with a CN in M81. We suppose that the most probable candidates are flares of M-type
stars. Based on a personal communication with S. Hawley (she used a model of Galactic M dwarf flare rates,
Hawley et al. 2007) we are expecting at most 1 flare brighter than $r' = 21.5$ mag of M-type star fainter
than $r' = 23.5$ mag during quiescent stage (see paragraph below), per square degree per year, in line
of sight of M81.

Thus, we estimate that we could have recorded about one flare in the interval of 90 months. Like in the case
of background SNe, gaps in time coverage by our images in order of months could only decrease the probability
of detection of such flare.

All objects that were classified as CN candidates have $R = 21.5$ mag or brighter on available images with
the exception of two objects (M81N~2004-09a and M81N~2005-01a). We discard all of M dwarfs brighter
than $r' = 23.5$ mag during quiescent stage because M dwarfs brighter than $r' = 23.5$ mag would be detected
in our deep images and classified as variable stars.

Although the uncertainty of estimate of number of flares of M-type stars is large (to a factor of a few),
mainly because an exact number of Galactic M dwarfs fainter than $r' = 22$ mag is unknown, we conclude that
the contamination of our sample of CN candidates by flares of M-type stars is very low or nonexistent.

We cannot be absolutely sure that all of our CN candidates are really CNe because of the limited data
available, we find that the contamination by other types of variable stars is probably very small. In any case,
the contamination did not significantly affect the results of the spatial distribution of CN candidates.

\subsection{Gaps in time coverage}\label{gaps}
As is mentioned above, the time spacing between images used differ, and long gaps were
present between images covering whole area of the galaxy.
Certainly, this would affect the number of CN candidates detected in the outer regions of the galaxy in comparison
with central part (bulge), where the gaps between images were smaller.

Moreover, if a hypothesis that fast novae of He/N spectroscopic class are abundant in the disc
population is right, then the number of CN candidates detected in the outer parts of the galaxy
could be even more affected. That is because, though the novae of He/N class have higher maximal
brightness, they typically have much more faster decline of luminosity than novae of FeII
spectroscopic class.

Since the CNe are expected, but did not detected, in the outer regions of the galaxy,
we simulated a probability that we did not detected any CN candidate there.
We used 27 images, from which 23 cover whole galaxy and the rest most of the area of the galaxy.
Most of them were obtained using narrow-band $H_{\alpha}$ filter. The limiting magnitude
of the images differ slightly, with an average value of 21.0 mag.
Using distance modulus of 27.8 mag for M81, this gives an absolute limiting magnitude in
$H_{\alpha}$ of $-6.8$ mag.

The simulation requires an estimate of the lifetime in $H_{\alpha}$ of a typical nova, which means
a number of days that a typical nova remains brighter than a given absolute
magnitude. Using an assumption that novae in M81 have the same characteristics as novae
in M31, we estimated the mean lifetime of a typical nova in $H_{\alpha}$ on an average image covering
the outer regions of the galaxy as $\sim 100$ days, or slightly more. This number was estimated from
lifetime relation shown in Fig. 13 of Shafter \& Irby (2001), where we used an absolute magnitude
in $H_{\alpha}$ of $-8.9$ mag, according to average $H_{\alpha}$ maximum magnitude for M31 novae
of $\sim 15.5$ mag (Ciardullo et al. 1990) and distance modulus of 24.4 mag. The above values
($-8.9$ mag and 15.5 mag) can be slightly lower (i.e., the brightnesses higher), since
Ciardullo et al. (1990) missed maximum light phase of some of their detected novae.

Next, we estimated the mean lifetime of novae of the He/N spectroscopic class. These novae are
typically brighter (we are assuming maximum absolute magnitude in $H_{\alpha}$ of $\sim -10$ mag),
but they have higher decline rate (assuming $\sim 0.03$ mag/day). Thus, we esimated their mean lifetime
as $\sim 100$ days too.

We assume that extremely fast novae are not very common; the assumption is based on results
published by Ciardullo et al. (1990), Shafter \& Irby (2001), and Neill \& Shara (2004). However,
we estimated their mean lifetime to be $\sim 40$ days by assuming that they can reach
an absolute magnitude of
$\sim -10.5$ in $H_{\alpha}$ and have a corresponding rate of decline of $\sim 0.10$ mag/day.

\begin{figure}[!h]
\includegraphics[width=0.48\textwidth]{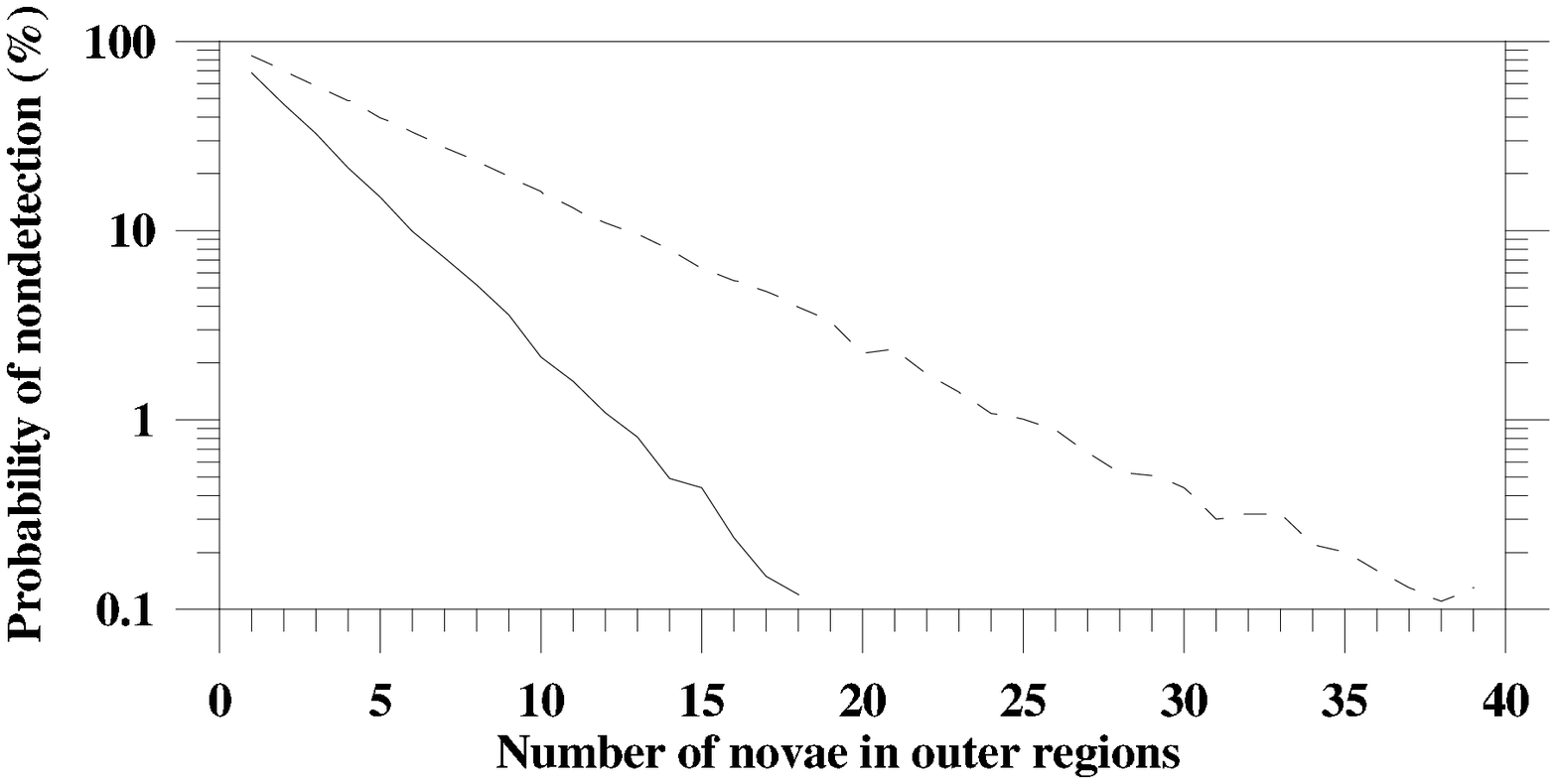}
\caption{Probability of nondetection of novae in our images covering the outer regions of the galaxy
depending on number of novae explosions which really occured. Novae with mean lifetime of 100 and 40 days
are plotted with solid and dotted lines, respectively.}
\label{9592fig8}
\end{figure}

In the simulation, we randomly distribute $K$ novae explosions in the time covered by all of 60 images
used in our work, which was equal to 2735 days. If some of the 27 images (from which 23 cover whole galaxy
and the rest most of the area of the galaxy) were taken during the nova lifetime, the nova was considered as detected.
The process was repeated 10000 times, and the probability of nondetection of any nova was calculated as
$N_0/10000$, where $N_0$ is number of runs during which no novae were detected on any of the 27 images.
The results for $K$ from 1 to 40 and lifetimes of 100 and 40 days are presented in Fig.~\ref{9592fig8}.
The probability of nondetection drops below 20\% for 5 and 9 novae with the lifetimes of 100 and 40 days,
respectively. Based on these numbers, we concluded that we cannot rule out the possibility that our lack
of novae detections in the outer regions of the galaxy is just a coincidence, and that we probably missed a
few novae there.

\section{Conclusions}
    We discovered and classified 49 transient objects as M81 CN candidates. Together with
previously known CN candidates (35 objects in total; published by Shara et al., 1999 and by
Neill \& Shara, 2004), the number of all CN candidates is more than doubled.
These results are important for future studies concerning the identification of optical
counterparts of supersoft X-ray sources in the M81 galaxy and for possible identification
of recurrent novae.

{The relatively large number of CN candidates in this sample 
provides a more accurate view
of the CN spatial distribution in M81 than we had before this study. There is no strong evidence
of asymmetry in the distribution of our CN candidates across the major axis of M81.
We cannot sustain the claim of a very high bulge-to-disk nova ratio in M81. Our results from the BTR
diagnostic have give a low bulge-to-disk nova ratio and thus support the existence of a significant
disk nova population.

The KS test applied to the radial distribution of our CN candidates and the M81 light distributions
indicate a good match between the CN candidates and the total light of the parent galaxy 
with a high confidence level.
This indicates that both CN populations, i.e., the disk and the bulge, are certainly
present in the galaxy.

However, since we cannot rule out the possibility that we have missed a few novae in the outer
part of the galaxy due to sparse time coverage with respect to the central part of the galaxy
(and thus we tested conformity of distribution of galaxy light and CN candidates only up to
radius of 323\arcs), the conclusions from this method are probably of lower significance than
the BTR statistics.

We are not able to refine the annual nova rate due to time gaps in the coverage and
a generally low cadence of images. However, the number of CN candidates detected is consistent
with the previous result of 30 yr$^{-1}$ (Neill \& Shara, 2004) within relatively large uncertainties.

We have shown that relatively inhomogeneous material, as well as use of archival data originally
obtained for different purposes, can be used to obtain significant results. However, only
a long-period (three years running at least), deep (going down to $H_{\alpha}$ = 21 mag) and
comprehensive survey covering the whole galaxy with minimum  detection incompleteness
is able to provide an accurate bulge-to-disk nova ratio. Such a survey would significantly refine
the characteristics of both bulge and disk nova populations as well as the total annual nova rate.

\medskip

\begin{acknowledgements}
This paper makes use of data obtained from the Isaac Newton Group Archive which is maintained as part
of the CASU Astronomical Data Centre at the Institute of Astronomy, Cambridge. Based [in part] on data
collected at [Subaru Telescope] and obtained from the SMOKA, which is operated by the Astronomy Data
Center, National Astronomical Observatory of Japan. Guest User, Canadian Astronomy Data Centre, which
is operated by the Herzberg Institute of Astrophysics, National Research Council of Canada.
We want to thank to anonymous referee for his comments and recommendations, which helped to
improve the manuscript considerably. We are grateful for spectroscopic confirmation of nova
candidate M81N~2007-04b to J. M. Silverman, R. J. Foley and A. V. Filippenko and for obtaining
and providing of M81 images to P. Caga\v{s}, V. P\v{r}ib\'{\i}k, P. Caga\v{s}, Jr., G. Sostero,
L. Donato, M. Gonano, V.  Gonano, A. Tonelli, A. Lepardo and V. Santini. We thanks to A. Karska
for assistance with getting of images with the WIYN telescope; to R.W. Argyle for help with
obtaining the data from the ING Archive; to P. Caga\v{s} for affording of SIMS software, to F. Hroch
for affording of Munipack software and to M.~Velen and P.~Pravec for affording of Aphot software.
Also we thanks to M.~Wolf and W.~Pietsch for valuable comments.

This research has made use of the SIMBAD database, operated at CDS,
Strasbourg, France, and of NASA's Astrophysics Data System Bibliographic
Services.
\end{acknowledgements}

\scriptsize
\begin{longtable}{llrrrr}
\caption{Photometry of nova candidates.
Observers, observatories, telescopes and CCDs for measurements are coded in column
comment according to Table~\ref{tab1}. Note that the measurements in F814W and F658N filters
of HST-ACS are in STMAG photometric system. Measurements in band coded as ``clear'' are from
unfiltered images, using R-band magnitudes from comparison stars. Measurements with errors in
order of hundredths of magnitude are stated in two decimal places; measurements with
errors $>$ 0.1 mag and $<$ 0.3 mag are stated in one decimal place; the ``:'' mark is used
for measurements with uncertainty $>$ 0.3 mag.}\label{tab4} \\
\hline\hline
JD& Mag & Band & Comment \\
(2\,450\,000+) \\
\hline
\endfirsthead
\caption{continued.}\\
\hline\hline
JD & Mag & Band & Comment \\
(2\,450\,000+) \\
\hline
\endhead
\hline
\endfoot
\noalign{\smallskip}
\multicolumn{6}{l}{Nova No. 1999-1 = \object{M81N~1999-11a}}\\
\noalign{\smallskip}
1501.982 &     21.21 & R   &   (13) \\
1501.998 &     21.5  & V   &   (13) \\
1877.998 &    [22.4  & V   &   (13) \\
1903.143 &    [23.6  & V   &   (16) \\
1912.626 &    [21.5  & Ha  &   (2) \\
2632.684 &    [22.7  & r   &   (3) \\
2632.721 &    [21.2  & Ha  &   (3) \\
2781.432 &    [21.6  & Ha  &   (4) \\
3378.917 &    [24.0  & V   &   (17) \\
3680.766 &    [23.8  & R   &   (19) \\
3792.677 &    [20.8  & Ha  &   (1) \\
3792.697 &    [21.4  & R   &   (1) \\
3793.869 &    [21.0  & Ha  &   (1) \\
\noalign{\smallskip}
\multicolumn{6}{l}{Nova No. 1999-2 = \object{M81N~1999-11b}}\\
\noalign{\smallskip}
1501.982 &     21.4  & R   &   (13) \\
1501.998 &     21.5  & V   &   (13) \\
1877.998 &    [22.4  & V   &   (13) \\
1903.143 &    [23.6  & V   &   (16) \\
1912.626 &    [21.5  & Ha  &   (2) \\
2632.684 &    [22.7  & r   &   (3) \\
2632.721 &    [21.2  & Ha  &   (3) \\
2781.432 &    [21.6  & Ha  &   (4) \\
3378.917 &    [24.0  & V   &   (17) \\
3680.766 &    [23.8  & R   &   (19) \\
3792.677 &    [20.8  & Ha  &   (1) \\
3792.697 &    [21.4  & R   &   (1) \\
3793.869 &    [21.0  & Ha  &   (1) \\
\noalign{\smallskip}
\multicolumn{6}{l}{Nova No. 2000-1 = \object{M81N~2000-11a}}\\
\noalign{\smallskip}
1501.982 &    [22.6  & R   &   (13) \\
1877.982 &     21.5  & R   &   (13) \\
1877.998 &    [22.4  & V   &   (13) \\
1903.143 &    [23.6  & V   &   (16) \\
1912.626 &    [21.5  & Ha  &   (2) \\
2632.684 &    [22.7  & r   &   (3) \\
2632.721 &    [21.2  & Ha  &   (3) \\
2781.432 &    [21.6  & Ha  &   (4) \\
3378.917 &    [24.0  & V   &   (17) \\
3680.766 &    [23.8  & R   &   (19) \\
3792.677 &    [20.8  & Ha  &   (1) \\
3792.697 &    [21.4  & R   &   (1) \\
3793.869 &    [21.0  & Ha  &   (1) \\
\noalign{\smallskip}
\multicolumn{6}{l}{Nova No. 2000-2 = \object{M81N~2000-12a}}\\
\noalign{\smallskip}
1501.982 &    [22.6  & R   &   (13) \\
1877.982 &    [22.5  & R   &   (13) \\
1877.998 &    [22.4  & V   &   (13) \\
1903.143 &     21.93 & V   &   (16) \\
1903.158 &     23.2  & B   &   (16) \\
1912.626 &     20.76 & Ha  &   (2) \\
2632.684 &    [22.7  & r   &   (3) \\
2632.721 &    [21.2  & Ha  &   (3) \\
2781.432 &    [21.6  & Ha  &   (4) \\
3378.917 &    [24.0  & V   &   (17) \\
3792.677 &    [20.8  & Ha  &   (1) \\
3792.697 &    [21.4  & R   &   (1) \\
3793.869 &    [21.0  & Ha  &   (1) \\
\noalign{\smallskip}
\multicolumn{6}{l}{Nova No. 2001-1 = \object{M81N~2001-01a}}\\
\noalign{\smallskip}
1903.143 &     21.37 & V   &   (16) \\
1903.158 &     21.93 & B   &   (16) \\
1912.626 &     18.32 & Ha  &   (2) \\
2632.721 &    [21.2  & Ha  &   (3) \\
2781.432 &    [21.6  & Ha  &   (4) \\
3378.917 &    [24.0  & V   &   (17) \\
3680.766 &    [23.8  & R   &   (19) \\
3792.677 &    [20.8  & Ha  &   (1) \\
3792.697 &    [21.4  & R   &   (1) \\
3793.869 &    [21.0  & Ha  &   (1) \\
\noalign{\smallskip}
\multicolumn{6}{l}{Nova No. 2001-2 = \object{M81N~2001-01b}}\\
\noalign{\smallskip}
1903.143 &     19.88 & V   &   (16) \\
1903.158 &     20.41 & B   &   (16) \\
1912.626 &     19.63 & Ha  &   (2) \\
2632.721 &    [21.2  & Ha  &   (3) \\
2781.432 &    [21.6  & Ha  &   (4) \\
3378.917 &    [24.0  & V   &   (17) \\
3680.766 &    [23.8  & R   &   (19) \\
3792.677 &    [20.8  & Ha  &   (1) \\
3792.697 &    [21.4  & R   &   (1) \\
3793.869 &    [21.0  & Ha  &   (1) \\
\noalign{\smallskip}
\multicolumn{6}{l}{Nova No. 2001-3 = \object{M81N~2001-01c}}\\
\noalign{\smallskip}
1903.143 &     23.0  & V   &   (16) \\
1903.158 &     23.1  & B   &   (16) \\
1912.626 &     20.3  & Ha  &   (2) \\
2632.721 &    [21.2  & Ha  &   (3) \\
2781.432 &    [21.6  & Ha  &   (4) \\
3378.917 &    [24.0  & V   &   (17) \\
3680.766 &    [23.8  & R   &   (19) \\
3792.677 &    [20.8  & Ha  &   (1) \\
3792.697 &    [21.4  & R   &   (1) \\
3793.869 &    [21.0  & Ha  &   (1) \\
\noalign{\smallskip}
\multicolumn{6}{l}{Nova No. 2001-4 = \object{M81N~2001-01d}}\\
\noalign{\smallskip}
1903.143 &     23.0  & V   &   (16) \\
1903.158 &     22.9  & B   &   (16) \\
1912.626 &     19.7  & Ha  &   (2) \\
2632.721 &    [21.2  & Ha  &   (3) \\
2781.432 &    [21.6  & Ha  &   (4) \\
3378.917 &    [24.0  & V   &   (17) \\
3680.766 &    [23.8  & R   &   (19) \\
3792.677 &    [20.8  & Ha  &   (1) \\
3792.697 &    [21.4  & R   &   (1) \\
3793.869 &    [21.0  & Ha  &   (1) \\
\noalign{\smallskip}
\multicolumn{6}{l}{Nova No. 2001-5 = \object{M81N~2001-01e}}\\
\noalign{\smallskip}
1877.982 &     21.1  & R   &   (13) \\
1877.998 &     21.3  & V   &   (13) \\
1912.626 &     19.9  & Ha  &   (2) \\
2632.721 &    [21.2  & Ha  &   (3) \\
2781.432 &    [21.6  & Ha  &   (4) \\
3680.766 &    [23.8  & R   &   (19) \\
3792.677 &    [20.8  & Ha  &   (1) \\
3792.697 &    [21.4  & R   &   (1) \\
3793.869 &    [21.0  & Ha  &   (1) \\
\noalign{\smallskip}
\multicolumn{6}{l}{Nova No. 2001-6 = \object{M81N~2001-01f}}\\
\noalign{\smallskip}
1903.143 &    [23.6  & V   &   (16) \\
1903.158 &    [24.1  & B   &   (16) \\
1912.626 &     20.32 & Ha  &   (2) \\
2632.721 &    [21.2  & Ha  &   (3) \\
2781.432 &    [21.6  & Ha  &   (4) \\
3680.766 &    [23.8  & R   &   (19) \\
3792.677 &    [20.8  & Ha  &   (1) \\
3792.697 &    [21.4  & R   &   (1) \\
3793.869 &    [21.0  & Ha  &   (1) \\
\noalign{\smallskip}
\multicolumn{6}{l}{Nova No. 2001-7 = \object{M81N~2001-01g}}\\
\noalign{\smallskip}
1903.143 &     22.0  & V   &   (16) \\
1903.158 &     23.1  & B   &   (16) \\
1912.626 &     21.06 & Ha  &   (2) \\
2632.721 &    [21.2  & Ha  &   (3) \\
2781.432 &    [21.6  & Ha  &   (4) \\
3378.928 &    [24.0  & V   &   (17) \\
3792.677 &    [20.8  & Ha  &   (1) \\
3792.697 &    [21.4  & R   &   (1) \\
3793.869 &    [21.0  & Ha  &   (1) \\
\noalign{\smallskip}
\multicolumn{6}{l}{Nova No. 2001-8 = \object{M81N~2001-01h}}\\
\noalign{\smallskip}
1903.143 &     23.2  & V   &   (16) \\
1903.158 &     23.7  & B   &   (16) \\
1912.626 &     20.63 & Ha  &   (2) \\
2632.721 &    [21.2  & Ha  &   (3) \\
2781.432 &    [21.6  & Ha  &   (4) \\
3378.928 &    [24.0  & V   &   (17) \\
3680.766 &    [23.8  & R   &   (19) \\
3792.677 &    [20.8  & Ha  &   (1) \\
3792.697 &    [21.4  & R   &   (1) \\
3793.869 &    [21.0  & Ha  &   (1) \\
\noalign{\smallskip}
\multicolumn{6}{l}{Nova No. 2002-1 = \object{M81N~2002-12a}}\\
\noalign{\smallskip}
1912.626 &    [21.5  & Ha   &  (2) \\
2632.684 &     19.39 & r    &  (3) \\
2632.721 &     19.86 & Ha   &  (3) \\
2781.432 &    [21.6  & Ha   &  (4) \\
3680.766 &    [23.8  & R    &  (19) \\
3792.677 &    [20.8  & Ha   &  (1) \\
3792.697 &    [21.4  & R    &  (1) \\
3793.869 &    [21.0  & Ha   &  (1) \\
\noalign{\smallskip}
\multicolumn{6}{l}{Nova No. 2002-2 = \object{M81N~2002-12b}}\\
\noalign{\smallskip}
1912.626 &    [20.9  & Ha   &  (2) \\
2632.684 &     20.8: & r    &  (3) \\
2632.721 &     18.9  & Ha   &  (3) \\
2781.432 &    [20.5  & Ha   &  (4) \\
3792.677 &    [20.8  & Ha   &  (1) \\
3792.697 &    [21.4  & R    &  (1) \\
3793.869 &    [21.0  & Ha   &  (1) \\
\noalign{\smallskip}
\multicolumn{6}{l}{Nova No. 2002-3 = \object{M81N~2002-12c}}\\
\noalign{\smallskip}
1912.626 &    [21.5  & Ha   &  (2) \\
2632.684 &    [22.3  & r    &  (3) \\
2632.721 &     21.05 & Ha   &  (3) \\
2781.432 &    [21.4  & Ha   &  (4) \\
3680.766 &    [23.8  & R    &  (19) \\
3792.677 &    [20.8  & Ha   &  (1) \\
3792.697 &    [21.4  & R    &  (1) \\
3793.869 &    [21.0  & Ha   &  (1) \\
\noalign{\smallskip}
\multicolumn{6}{l}{Nova No. 2002-4 = \object{M81N~2002-12d}}\\
\noalign{\smallskip}
1912.626 &    [21.5  & Ha   &  (2) \\
2632.684 &     21.10 & r    &  (3) \\
2632.721 &     19.72 & Ha   &  (3) \\
2781.432 &    [21.6  & Ha   &  (4) \\
3680.766 &    [23.8  & R    &  (19) \\
3792.677 &    [20.8  & Ha   &  (1) \\
3792.697 &    [21.4  & R    &  (1) \\
3793.869 &    [21.0  & Ha   &  (1) \\
\noalign{\smallskip}
\multicolumn{6}{l}{Nova No. 2002-5 = \object{M81N~2002-12e}}\\
\noalign{\smallskip}
1912.626 &    [21.5  & Ha   &  (2) \\
2617.039 &     20.52 & R    &  (13) \\
2632.684 &     21.42 & r    &  (3) \\
2632.721 &     19.82 & Ha   &  (3) \\
2781.432 &    [21.6  & Ha   &  (4) \\
3680.766 &    [23.8  & R    &  (19) \\
3792.677 &    [20.8  & Ha   &  (1) \\
3792.697 &    [21.4  & R    &  (1) \\
3793.869 &    [21.0  & Ha   &  (1) \\
\noalign{\smallskip}
\multicolumn{6}{l}{Nova No. 2003-1 = \object{M81N~2003-05c}}\\
\noalign{\smallskip}
1912.626 &    [21.5  & Ha   &  (2) \\
2769.383 &     21.8  & r    &  (7) \\
2769.388 &     22.0: & Ha   &  (7) \\
2781.432 &     19.70 & Ha   &  (4) \\
2781.436 &     20.61 & r    &  (4) \\
2783.455 &     19.59 & Ha   &  (4) \\
2783.460 &     20.39 & r    &  (4) \\
2785.450 &     19.71 & Ha   &  (4) \\
2794.391 &     19.10 & Ha   &  (5) \\
2794.397 &     20.73 & r    &  (5) \\
2795.394 &     19.03 & Ha   &  (5) \\
2796.395 &     18.91 & Ha   &  (5) \\
2796.400 &     20.59 & r    &  (5) \\
2800.384 &     19.05 & Ha   &  (6) \\
3680.766 &    [23.8  & R    &  (19) \\
3792.677 &    [20.8  & Ha   &  (1) \\
3793.869 &    [21.0  & Ha   &  (1) \\
\noalign{\smallskip}
\multicolumn{6}{l}{Nova No. 2003-2 = \object{M81N~2003-09a}}\\
\noalign{\smallskip}
1912.626 &    [21.3  & Ha   &  (2) \\
2800.384 &    [21.4  & Ha   &  (6) \\
2900.882 &     22.34 & F814W &  (9) \\
2900.895 &     20.24 & F658N &  (9) \\
3039.528 &    [22.3  & V    &  (7) \\
3039.540 &     20.10 & Ha   &  (7) \\
3040.771 &    [20.0  & Ha   &  (7) \\
3049.631 &     21.0: & Ha   &  (8) \\
3049.660 &    [22.0  & V    &  (8) \\
3052.735 &    [22.3  & V    &  (8) \\
3052.744 &     20.6: & Ha   &  (8) \\
3263.850 &    [25.3  & F814W &  (11) \\
3680.766 &    [23.8  & R    &  (19) \\
3792.677 &    [20.8  & Ha   &  (1) \\
3793.869 &    [21.0  & Ha   &  (1) \\
\noalign{\smallskip}
\multicolumn{6}{l}{Nova No. 2003-3 = \object{M81N~2003-09b}}\\
\noalign{\smallskip}
1912.626 &    [21.3  & Ha   &  (2) \\
2422.782 &    [27.3  & F814W &  (10) \\
2800.384 &    [21.4  & Ha   & (6) \\
2900.882 &     23.28 & F814W &  (9) \\
2900.895 &     20.38 & F658N &  (9) \\
3039.528 &    [22.3  & V     & (7) \\
3039.540 &    [21.1  & Ha    & (7) \\
3263.780 &    [25.4  & F814W &  (11) \\
3792.677 &    [20.8  & Ha    & (1) \\
3793.869 &    [21.0  & Ha    & (1) \\
\noalign{\smallskip}
\multicolumn{6}{l}{Nova No. 2003-4 = \object{M81N~2003-09c}}\\
\noalign{\smallskip}
1912.626 &    [21.0  & Ha    & (2) \\
2800.384 &    [21.1  & Ha    & (6) \\
2900.882 &     23.7  & F814W &  (9) \\
2900.895 &     19.45 & F658N &  (9) \\
3039.528 &    [22.0  & V     & (7) \\
3039.540 &    [20.8  & Ha    & (7) \\
3263.850 &    [25.4  & F814W &  (11) \\
3680.766 &    [23.8  & R     & (19) \\
3792.677 &    [20.8  & Ha    & (1) \\
3793.869 &    [21.0  & Ha    & (1) \\
\noalign{\smallskip}
\multicolumn{6}{l}{Nova No. 2003-5 = \object{M81N~2003-05a}}\\
\noalign{\smallskip}
1912.626 &    [21.5  & Ha   &  (2) \\
2632.684 &    [22.7  & r    &  (3) \\
2632.721 &    [21.2  & Ha   &  (3) \\
2769.383 &     20.26 & r    &  (7) \\
2769.388 &     18.72 & Ha   &  (7) \\
2781.432 &     20.63 & Ha   &  (4) \\
2781.436 &    [22.4  & r    &  (4) \\
2783.455 &     20.58 & Ha   &  (4) \\
2783.460 &    [21.5  & r    &  (4) \\
2785.450 &     21.0  & Ha   &  (4) \\
2794.391 &    [20.9  & Ha   &  (5) \\
2794.397 &    [21.1  & r    &  (5) \\
2795.394 &    [21.4  & Ha   &  (5) \\
2796.395 &     22.4: & Ha   &  (5) \\
2800.384 &    [21.4  & Ha   &  (6) \\
2900.895 &    [23.0  & F658N &  (9) \\
3039.540 &    [21.5  & Ha   &  (7) \\
3680.766 &    [23.8  & R    &  (19) \\
3792.677 &    [20.8  & Ha   &  (1) \\
3793.869 &    [21.0  & Ha   &  (1) \\
\noalign{\smallskip}
\multicolumn{6}{l}{Nova No. 2003-6 = \object{M81N~2003-05b}}\\
\noalign{\smallskip}
1912.626 &    [21.5  & Ha   &  (2) \\
2632.684 &    [22.7  & r    &  (3) \\
2632.721 &    [21.2  & Ha   &  (3) \\
2769.383 &     21.4  & r    &  (7) \\
2769.388 &     19.64 & Ha   &  (7) \\
2781.432 &     19.85 & Ha   &  (4) \\
2781.436 &     21.3  & r    &  (4) \\
2783.455 &     19.84 & Ha   &  (4) \\
2783.460 &     21.5  & r    &  (4) \\
2785.450 &     19.83 & Ha   &  (4) \\
2794.391 &     20.0  & Ha   &  (5) \\
2794.397 &    [21.1  & r    &  (5) \\
2795.394 &     20.3  & Ha   &  (5) \\
2796.395 &     20.6  & Ha   &  (5) \\
2800.384 &     20.4  & Ha   &  (6) \\
3039.540 &    [21.5  & Ha   &  (7) \\
3792.677 &     [20.8  & Ha   &  (1) \\
3793.869 &     [21.0  & Ha   &  (1) \\
\noalign{\smallskip}
\multicolumn{6}{l}{Nova No. 2004-1 = \object{M81N~2004-02c}}\\
\noalign{\smallskip}
1912.626 &    [21.5  & Ha   &  (2) \\
3039.528 &    [22.3  & V    &  (7) \\
3039.540 &    [21.1  & Ha   &  (7) \\
3040.765 &    [21.7  & V    &  (7) \\
3040.771 &    [20.0  & Ha   &  (7) \\
3040.781 &    [21.5  & r    &  (7) \\
3049.631 &     19.41 & Ha   &  (8) \\
3049.660 &     20.75 & V    &  (8) \\
3052.735 &    [21.8  & V    &  (8) \\
3052.744 &     19.20 & Ha   &  (8) \\
3680.766 &    [23.8  & R    &  (19) \\
3792.677 &    [20.8  & Ha   &  (1) \\
3793.869 &    [21.0  & Ha   &  (1) \\
\noalign{\smallskip}
\multicolumn{6}{l}{Nova No. 2004-2 = \object{M81N~2004-02a}}\\
\noalign{\smallskip}
1912.626 &    [21.5  & Ha   &  (2) \\
3039.528 &     22.1  & V    &  (7) \\
3039.540 &     20.47 & Ha   &  (7) \\
3040.765 &    [21.7  & V    &  (7) \\
3040.771 &    [20.0  & Ha   &  (7) \\
3040.781 &    [21.5  & r    &  (7) \\
3049.631 &     20.8  & Ha   &  (8) \\
3049.660 &    [22.0  & V    &  (8) \\
3052.735 &    [22.2  & V    &  (8) \\
3052.744 &     20.7: & Ha   &  (8) \\
3680.766 &    [23.8  & R    &  (19) \\
3792.677 &    [20.8  & Ha   &  (1) \\
3793.869 &    [21.0  & Ha   &  (1) \\
\noalign{\smallskip}
\multicolumn{6}{l}{Nova No. 2004-3 = \object{M81N~2004-02b}}\\
\noalign{\smallskip}
1912.626 &    [21.5  & Ha   &  (2) \\
3039.528 &     21.14 & V    &  (7) \\
3039.540 &     19.92 & Ha   &  (7) \\
3049.631 &     19.69 & Ha   &  (8) \\
3049.660 &     21.6  & V    &  (8) \\
3052.735 &     21.7  & V    &  (8) \\
3052.744 &     19.58 & Ha   &  (8) \\
3079.781 &     21.30 & R    &  (13) \\
3080.861 &     21.40 & R    &  (12) \\
3792.677 &    [20.8  & Ha   &  (1) \\
3793.869 &    [21.0  & Ha   &  (1) \\
\noalign{\smallskip}
\multicolumn{6}{l}{Nova No. 2004-4 = \object{M81N~2004-09a}}\\
\noalign{\smallskip}
1912.626 &    [21.3  & Ha   &  (2) \\
2632.684 &    [22.7  & r    &  (3) \\
2632.721 &    [21.2  & Ha   &  (3) \\
2781.432 &    [21.6  & Ha   &  (4) \\
2900.882 &    [25.9  & F814W &   (9) \\
2900.895 &    [23.0  & F658N &  (9) \\
3039.528 &    [22.3  & V    &  (7) \\
3039.540 &    [20.8  & Ha   &  (7) \\
3052.744 &    [21.1  & Ha   &  (8) \\
3263.280 &     22.62 & F814W &  (11) \\
3680.766 &    [23.8  & R    &  (19) \\
3792.677 &    [20.8  & Ha   &  (1) \\
3793.869 &    [21.0  & Ha   &  (1) \\
\noalign{\smallskip}
\multicolumn{6}{l}{Nova No. 2005-1 = \object{M81N~2005-12a}}\\
\noalign{\smallskip}
1912.626 &    [21.3  & Ha   &  (2) \\
2632.684 &    [22.7  & r    &  (3) \\
2781.432 &    [21.6  & Ha   &  (4) \\
3052.744 &    [21.1  & Ha   &  (8) \\
3712.776 &     20.2  & Ha   &  (14) \\
3793.865 &    [22.5  & R    &  (1) \\
3793.869 &    [21.0  & Ha   &  (1) \\
4081.914 &    [21.6  & R    &  (15) \\
4081.939 &    [19.9  & R    &  (15) \\
\noalign{\smallskip}
\multicolumn{6}{l}{Nova No. 2005-2 = \object{M81N~2005-12b}}\\
\noalign{\smallskip}
1912.626 &    [21.3  & Ha   &  (2) \\
2632.684 &    [22.7  & r    &  (3) \\
2781.432 &    [21.6  & Ha   &  (4) \\
3052.744 &    [21.1  & Ha   &  (8) \\
3680.766 &    [23.8  & R    &  (19) \\
3712.776 &     20.1  & Ha   &  (14) \\
3750.987 &    [23.1  & V    &  (31) \\
3776.062 &    [22.5  & Ha   &  (31) \\
3793.865 &    [22.5  & R    &  (1) \\
3793.869 &    [21.0  & Ha   &  (1) \\
4081.914 &    [21.6  & R    &  (15) \\
4081.939 &    [19.9  & Ha   &  (15) \\
\noalign{\smallskip}
\multicolumn{6}{l}{Nova No. 2005-3 = \object{M81N~2005-01a}}\\
\noalign{\smallskip}
1903.143 &    [23.6  & V    &  (16) \\
1912.626 &    [21.5  & Ha   &  (2) \\
2781.432 &    [21.6  & Ha   &  (4) \\
3052.744 &    [21.1  & Ha   &  (8) \\
3377.951 &     22.5  & V    &  (17) \\
3378.928 &     22.3  & V    &  (17) \\
3680.766 &    [23.8  & R    &  (19) \\
3793.865 &    [22.5  & R    &  (1) \\
3793.869 &    [21.0  & Ha   &  (1) \\
4081.914 &    [21.6  & R    &  (15) \\
4081.939 &    [19.9  & Ha   &  (15) \\
\noalign{\smallskip}
\multicolumn{6}{l}{Nova No. 2005-4 = \object{M81N~2005-11a}}\\
\noalign{\smallskip}
1912.626 &    [21.5  & Ha   &  (2) \\
2632.684 &    [22.7  & r    &  (3) \\
2781.432 &    [21.6  & Ha   &  (4) \\
3052.744 &    [21.1  & Ha   &  (8) \\
3680.766 &      21.1 & R    &  (19) \\
3712.776 &    [20.9  & Ha   &  (14) \\
3750.987 &    [23.1  & V    &  (31) \\
3776.062 &     22.2: & Ha   &  (31) \\
3793.865 &    [22.5  & R    &  (1) \\
3793.869 &    [21.0  & Ha   &  (1) \\
4081.914 &    [21.2  & R    &  (15) \\
4081.939 &    [20.3  & Ha   &  (15) \\
4124.767 &    [21.6  & Ha   &  (18) \\
4124.802 &    [22.0  & R    &  (18) \\
\noalign{\smallskip}
\multicolumn{6}{l}{Nova No. 2006-1 = \object{M81N~2006-02a}}\\
\noalign{\smallskip}
1912.626 &    [21.3  & Ha   &  (2) \\
2632.721 &    [21.2  & Ha   &  (3) \\
2781.432 &    [21.6  & Ha   &  (4) \\
3680.766 &    [23.8  & R    &  (19) \\
3750.987 &    [23.1  & V    &  (31) \\
3776.062 &    [22.5  & Ha   &  (31) \\
3792.677 &     19.2  & Ha   &  (1) \\
3792.697 &    [21.4  & R    &  (1) \\
3809.413 &    [20.9  & r    &  (20) \\
3809.395 &     20.6  & Ha   &  (20) \\
3813.457 &     23.0: & B    &  (21) \\
3814.490 &    [22.5  & B    &  (21) \\
3815.515 &    [22.0  & B    &  (21) \\
4124.767 &    [21.6  & Ha   &  (18) \\
\noalign{\smallskip}
\multicolumn{6}{l}{Nova No. 2006-2 = \object{M81N~2006-02b}}\\
\noalign{\smallskip}
1912.626 &    [21.0  & Ha   &  (2) \\
2632.721 &    [21.0  & Ha   &  (3) \\
2781.432 &    [21.4  & Ha   &  (4) \\
3680.766 &    [23.8  & R    &  (19) \\
3750.987 &    [23.1  & V    &  (31) \\
3776.062 &    [22.5  & Ha   &  (31) \\
3792.677 &     19.0  & Ha   &  (1) \\
3792.697 &    [21.4  & R    &  (1) \\
3793.865 &    [21.5  & R    &  (1) \\
3793.869 &     19.4  & Ha   &  (1) \\
3809.413 &    [20.9  & r    &  (20) \\
3809.395 &    [21.0  & Ha   &  (20) \\
3813.457 &    [23.0  & B    &  (21) \\
3814.490 &    [22.5  & B    &  (21) \\
3815.515 &    [22.0  & B    &  (21) \\
4124.767 &    [21.6  & Ha   &  (18) \\
\noalign{\smallskip}
\multicolumn{6}{l}{Nova No. 2006-3 = \object{M81N~2006-02c}}\\
\noalign{\smallskip}
1912.626 &    [21.3  & Ha   &  (2) \\
2632.721 &    [21.1  & Ha   &  (3) \\
2781.432 &    [21.5  & Ha   &  (4) \\
3680.766 &    [23.8  & R    &  (19) \\
3750.987 &    [23.1  & V    &  (31) \\
3776.062 &    [22.5  & Ha   &  (31) \\
3792.677 &     18.65 & Ha   &  (1) \\
3809.413 &    [21.0  & r    &  (20) \\
3809.395 &    [21.2  & Ha   &  (20) \\
3813.457 &    [23.5  & B    &  (21) \\
3814.490 &    [22.8  & B    &  (21) \\
3815.515 &    [22.2  & B    &  (21) \\
4124.767 &    [21.6  & Ha   &  (18) \\
\noalign{\smallskip}
\multicolumn{6}{l}{Nova No. 2006-4 = \object{M81N~2006-02d}}\\
\noalign{\smallskip}
1912.626 &    [21.3  & Ha   &  (2) \\
2632.721 &    [21.2  & Ha   &  (3) \\
2781.432 &    [21.6  & Ha   &  (4) \\
3680.766 &    [23.8  & R    &  (19) \\
3750.987 &    [23.1  & V    &  (31) \\
3776.062 &    [22.5  & Ha   &  (31) \\
3792.677 &     19.7  & Ha   &  (1) \\
3792.697 &     21.0: & R    &  (1) \\
3793.865 &    [21.3  & R    &  (1) \\
3793.869 &     19.9  & Ha   &  (1) \\
3809.413 &    [20.8  & r    &  (20) \\
3809.395 &     19.55 & Ha   &  (20) \\
3813.457 &     22.0  & B    &  (21) \\
3814.490 &     22.0  & B    &  (21) \\
3815.515 &     21.9  & B    &  (21) \\
4124.767 &    [21.6  & Ha   &  (18) \\
\noalign{\smallskip}
\multicolumn{6}{l}{Nova No. 2006-5 = \object{M81N~2006-02f}}\\
\noalign{\smallskip}
1912.626 &    [21.0  & Ha   &  (2) \\
2632.721 &    [21.2  & Ha   &  (3) \\
2781.432 &    [21.6  & Ha   &  (4) \\
3750.987 &    [23.1  & V    &  (31) \\
3776.062 &    [22.5  & Ha   &  (31) \\
3792.677 &     20.5: & Ha   &  (1) \\
3792.697 &     20.2  & R    &  (1) \\
3793.865 &     20.0  & R    &  (1) \\
3793.869 &     20.0  & Ha   &  (1) \\
3809.413 &    [20.9  & r    &  (20) \\
3809.395 &     19.7  & Ha   &  (20) \\
3813.457 &    [22.8  & B    &  (21) \\
3814.490 &    [22.5  & B    &  (21) \\
3815.515 &    [22.0  & B    &  (21) \\
4124.767 &    [21.6  & Ha   &  (18) \\
\noalign{\smallskip}
\multicolumn{6}{l}{Nova No. 2006-6 = \object{M81N~2006-02e}}\\
\noalign{\smallskip}
1912.626 &    [21.5  & Ha   &  (2) \\
2632.721 &    [21.2  & Ha   &  (3) \\
2781.432 &    [21.6  & Ha   &  (4) \\
3703.719 &     20.47 & R    &  (13) \\
3709.954 &     20.68 & R    &  (13) \\
3712.776 &     18.56 & Ha   &  (14) \\
3792.677 &     20.60 & Ha   &  (1) \\
3792.697 &     22.8: & R    &  (1) \\
3793.865 &     22.7  & R    &  (1) \\
3793.869 &     20.62 & Ha   &  (1) \\
3809.413 &    [21.0  & r    &  (20) \\
3809.395 &     21.0  & Ha   &  (20) \\
3813.457 &     23.5  & B    &  (21) \\
3814.490 &    [22.8  & B    &  (21) \\
3815.515 &    [22.4  & B    &  (21) \\
4124.767 &    [21.6  & Ha   &  (18) \\
\noalign{\smallskip}
\multicolumn{6}{l}{Nova No. 2006-7 = \object{M81N~2006-12a}}\\
\noalign{\smallskip}
1912.626 &    [21.3  & Ha   &  (2) \\
2632.684 &    [22.7  & r    &  (3) \\
2781.432 &    [21.6  & Ha   &  (4) \\
3052.744 &    [21.1  & Ha   &  (8) \\
3680.766 &    [23.8  & R    &  (19) \\
3712.776 &    [20.9  & Ha   &  (14) \\
3793.865 &    [22.5  & R    &  (1) \\
3793.869 &    [21.0  & Ha   &  (1) \\
4081.914 &     20.7  & R    &  (15) \\
4081.939 &     19.0  & Ha   &  (15) \\
4124.767 &    [21.6  & Ha   &  (18) \\
4124.802 &    [22.0  & R    &  (18) \\
\noalign{\smallskip}
\multicolumn{6}{l}{Nova No. 2006-8 = \object{M81N~2006-12b}}\\
\noalign{\smallskip}
1912.626 &    [21.5  & Ha   &   (2) \\
2632.684 &    [22.7  & r    &   (3) \\
2781.432 &    [21.6  & Ha   &   (4) \\
3052.744 &    [21.1  & Ha   &   (8) \\
3680.766 &    [23.8  & R    &   (19) \\
3712.776 &    [20.9  & Ha   &   (14) \\
3793.865 &    [22.5  & R    &   (1) \\
3793.869 &    [21.0  & Ha   &   (1) \\
4081.914 &     21.5: & R    &   (15) \\
4081.939 &     19.7  & Ha   &   (15) \\
4124.767 &     21.6: & Ha   &   (18) \\
4124.802 &    [22.0  & R    &   (18) \\
\noalign{\smallskip}
\multicolumn{6}{l}{Nova No. 2006-9 = \object{M81N~2006-03a}}\\
\noalign{\smallskip}
1903.158 &    [24.1  & B    &  (16) \\
1912.626 &    [21.5  & Ha   &  (2) \\
2632.684 &    [22.7  & r    &  (3) \\
2781.432 &    [21.6  & Ha   &  (4) \\
3052.744 &    [21.1  & Ha   &  (8) \\
3680.766 &    [23.8  & R    &  (19) \\
3712.776 &    [20.9  & Ha   &  (14) \\
3750.987 &    [23.1  & V    &  (31) \\
3776.062 &    [22.5  & Ha   &  (31) \\
3793.865 &    [22.8  & R    &  (1) \\
3793.869 &    [21.6  & Ha   &  (1) \\
3809.413 &     19.8  & r    &  (20) \\
3809.395 &     19.99 & Ha   &  (20) \\
3813.457 &     21.3  & B    &  (21) \\
3814.490 &     21.5  & B    &  (21) \\
3815.515 &     22.0  & B    &  (21) \\
4081.914 &    [21.2  & R    &  (15) \\
4081.939 &    [20.5  & Ha   &  (15) \\
4124.767 &    [21.6  & Ha   &  (18) \\
4124.802 &    [22.2  & R    &  (18) \\
\noalign{\smallskip}
\multicolumn{6}{l}{Nova No. 2006-10 = \object{M81N~2006-03b}}\\
\noalign{\smallskip}
1903.158 &    [24.1  & B    &  (16) \\
1912.626 &    [21.5  & Ha   &  (2) \\
2632.684 &    [22.7  & r    &  (3) \\
2781.432 &    [21.6  & Ha   &  (4) \\
3052.744 &    [21.1  & Ha   &  (8) \\
3680.766 &    [23.8  & R    &  (19) \\
3712.776 &    [20.9  & Ha   &  (14) \\
3750.987 &    [23.1  & V    &  (31) \\
3776.062 &    [22.5  & Ha   &  (31) \\
3793.865 &    [22.8  & R    &  (1) \\
3793.869 &    [21.6  & Ha   &  (1) \\
3809.413 &     20.4  & r    &  (20) \\
3809.395 &     20.66 & Ha   &  (20) \\
3813.457 &     20.79 & B    &  (21) \\
3814.490 &     21.06 & B    &  (21) \\
3815.515 &     21.3  & B    &  (21) \\
4081.914 &    [21.2  & R    &  (15) \\
4081.939 &    [20.5  & Ha   &  (15) \\
4124.767 &    [21.6  & Ha   &  (18) \\
4124.802 &    [22.2  & R    &  (18) \\
\noalign{\smallskip}
\multicolumn{6}{l}{Nova No. 2006-11 = \object{M81N~2006-02g}}\\
\noalign{\smallskip}
1903.158 &    [24.1  & B    &  (16) \\
1912.626 &    [21.5  & Ha   &  (2) \\
2632.684 &    [22.7  & r    &  (3) \\
2781.432 &    [21.6  & Ha   &  (4) \\
3052.744 &    [21.1  & Ha   &  (8) \\
3680.766 &    [23.8  & R    &  (19) \\
3712.776 &    [20.9  & Ha   &  (14) \\
3750.987 &     22.9: & V    &  (31) \\
3776.062 &     20.1  & Ha   &  (31) \\
3792.677 &     20.9  & Ha   &  (1) \\
3792.697 &    [22.8  & R    &  (1) \\
3793.865 &    [22.8  & R    &  (1) \\
3793.869 &     20.8  & Ha   &  (1) \\
3809.413 &    [21.0  & r    &  (20) \\
3809.395 &     20.9  & Ha   &  (20) \\
3813.457 &    [23.5  & B    &  (21) \\
3814.490 &    [22.8  & B    &  (21) \\
3815.515 &    [22.4  & B    &  (21) \\
4081.914 &    [21.2  & R    &  (15) \\
4081.939 &    [20.5  & Ha   &  (15) \\
4124.767 &    [21.6  & Ha   &  (18) \\
4124.802 &    [22.2  & R    &  (18) \\
\noalign{\smallskip}
\multicolumn{6}{l}{Nova No. 2006-12 = \object{M81N~2006-01a}}\\
\noalign{\smallskip}
1501.982 &    [22.6  & R    &  (13) \\
1903.143 &    [23.6  & V    &  (16) \\
1912.626 &    [21.5  & Ha   &  (2) \\
2632.684 &    [22.7  & r    &  (3) \\
2781.432 &    [21.6  & Ha   &  (4) \\
2900.895 &    [23.0  & F658N &  (9) \\
3379.017 &    [24.0  & V    &  (17) \\
3703.719 &    [22.3  & R    &  (13) \\
3709.954 &     21.8  & R    &  (13) \\
3750.987 &     20.9  & V    &  (31) \\
3776.062 &    [22.5  & Ha   &  (31) \\
3792.677 &    [20.8  & Ha   &  (1) \\
3792.697 &    [21.4  & R    &  (1) \\
3793.869 &    [21.0  & Ha   &  (1) \\
3813.457 &    [23.5  & B    &  (21) \\
\noalign{\smallskip}
\multicolumn{6}{l}{Nova No. 2006-13 = \object{M81N~2006-02h}}\\
\noalign{\smallskip}
1501.982 &    [22.6  & R    &  (13) \\
1912.626 &    [21.5  & Ha   &  (2) \\
2632.684 &    [22.7  & r    &  (3) \\
2781.432 &    [21.6  & Ha   &  (4) \\
2900.895 &    [23.0  & F658N & (9) \\
3703.719 &    [22.1  & R    &  (13) \\
3709.954 &    [22.1  & R    &  (13) \\
3750.987 &    [22.8  & V    &  (31) \\
3776.062 &     19.9  & Ha   &  (31) \\
3792.677 &    [20.8  & Ha   &  (1) \\
3792.697 &    [21.4  & R    &  (1) \\
3809.395 &    [20.8  & Ha   &  (20) \\
3813.457 &    [23.3  & B    &  (21) \\
4124.767 &    [21.6  & Ha   &  (18) \\
\noalign{\smallskip}
\multicolumn{6}{l}{Nova No. 2006-14 = \object{M81N~2006-02i}}\\
\noalign{\smallskip}
1501.982 &    [22.6  & R    &  (13) \\
1912.626 &    [21.5  & Ha   &  (2) \\
2632.684 &    [22.7  & r    &  (3) \\
2781.432 &    [21.6  & Ha   &  (4) \\
2900.895 &    [23.0  & F658N & (9) \\
3703.719 &    [22.3  & R    &  (13) \\
3709.954 &    [22.3  & R    &  (13) \\
3712.776 &    [20.9  & Ha   &  (14) \\
3750.987 &    [23.1  & V    &  (31) \\
3776.062 &     20.2  & Ha   &  (31) \\
3792.677 &     20.1  & Ha   &  (1) \\
3792.697 &    [21.4  & R    &  (1) \\
3809.395 &     21.1  & Ha   &  (20) \\
3813.457 &    [23.5  & B    &  (21) \\
4124.767 &    [21.6  & Ha   &  (18) \\
\noalign{\smallskip}
\multicolumn{6}{l}{Nova No. 2006-15 = \object{M81N~2006-02j}}\\
\noalign{\smallskip}
1501.982 &    [22.6  & R    &  (13) \\
1903.143 &    [23.6  & V    &  (16) \\
1912.626 &    [21.5  & Ha   &  (2) \\
2632.684 &    [22.7  & r    &  (3) \\
2781.432 &    [21.6  & Ha   &  (4) \\
2900.895 &    [23.0  & F658N & (9) \\
3703.719 &    [22.3  & R    &  (13) \\
3709.954 &    [22.3  & R    &  (13) \\
3712.776 &    [20.8  & Ha   &  (14) \\
3750.987 &     22.5: & V    &  (31) \\
3776.062 &     20.6  & Ha   &  (31) \\
3792.677 &     20.5  & Ha   &  (1) \\
3792.697 &    [21.4  & R    &  (1) \\
3793.865 &    [22.8  & R    &  (1) \\
3793.869 &     20.6  & Ha   &  (1) \\
3809.395 &     21.0  & Ha   &  (20) \\
3813.457 &    [23.5  & B    &  (21) \\
4124.767 &    [21.6  & Ha   &  (18) \\
\noalign{\smallskip}
\multicolumn{6}{l}{Nova No. 2007-1 = \object{M81N~2007-01a}}\\
\noalign{\smallskip}
1912.626 &    [21.5  & Ha   &  (2) \\
2632.684 &    [22.7  & r    &  (3) \\
2781.432 &    [21.6  & Ha   &  (4) \\
3052.744 &    [21.1  & Ha   &  (8) \\
3680.766 &    [23.8  & R    &  (19) \\
3712.776 &    [20.9  & Ha   &  (14) \\
3793.865 &    [22.5  & R    &  (1) \\
3793.869 &    [21.0  & Ha   &  (1) \\
4081.914 &    [21.2  & R    &  (15) \\
4081.939 &     19.6  & Ha   &  (15) \\
4124.767 &     20.9  & Ha   &  (18) \\
4124.802 &    [21.9  & R    &  (18) \\
\noalign{\smallskip}
\multicolumn{6}{l}{Nova No. 2007-2 = \object{M81N~2007-04a}}\\
\noalign{\smallskip}
1912.626 &    [21.0  & Ha   &  (2) \\
2632.684 &    [22.7  & r    &  (3) \\
2632.721 &    [21.0  & Ha   &  (3) \\
2781.432 &    [21.4  & Ha   &  (4) \\
3680.766 &    [22.9  & R    &  (19) \\
3792.677 &    [20.8  & Ha   &  (1) \\
3792.697 &    [21.4  & R    &  (1) \\
3793.865 &    [21.5  & R    &  (1) \\
3793.869 &    [21.0  & Ha   &  (1) \\
3809.413 &    [20.9  & r    &  (20) \\
3809.395 &    [21.0  & Ha   &  (20) \\
3813.457 &    [23.0  & B    &  (21) \\
3814.490 &    [22.5  & B    &  (21) \\
3815.515 &    [22.0  & B    &  (21) \\
4124.767 &    [21.6  & Ha   &  (18) \\
4124.802 &    [21.9  & R    &  (18) \\
4164.468 &    [20.5  & clear  &  (22) \\
4196.312 &     19.0: & clear  &  (22) \\
4199.374 &     19.2  & clear  &  (25) \\
4202.365 &     19.5  & clear  &  (24) \\
4203.409 &     19.4  & clear  &  (27) \\
4204.432 &     20.0: & clear  &  (25) \\
4205.358 &     20.2: & clear  &  (24) \\
\noalign{\smallskip}
\multicolumn{6}{l}{Nova No. 2007-3 = \object{M81N~2007-04b}}\\
\noalign{\smallskip}
1912.626 &    [21.4  & Ha    & (2) \\
2632.684 &    [22.7  & r     & (3) \\
2632.721 &    [21.0  & Ha    & (3) \\
2781.432 &    [21.6  & Ha    & (4) \\
3378.917 &    [24.0  & V     & (17) \\
3680.766 &    [23.0  & R     & (19) \\
3792.677 &    [20.8  & Ha    & (1) \\
3792.697 &    [21.4  & R     & (1) \\
3793.865 &    [22.5  & R     & (1) \\
3793.869 &    [21.0  & Ha    & (1) \\
3809.413 &    [20.9  & r     & (20) \\
3809.395 &    [21.0  & Ha    & (20) \\
3813.457 &    [23.0  & B     & (21) \\
3814.490 &    [22.5  & B     & (21) \\
3815.515 &    [22.0  & B     & (21) \\
4124.767 &    [21.6  & Ha    & (18) \\
4124.802 &    [22.0  & R     & (18) \\
4164.468 &    [21.5  & clear &  (22) \\
4196.312 &    [21.3  & clear &  (22) \\
4199.374 &    [21.0  & clear &  (25) \\
4202.365 &     19.9  & clear &  (24) \\
4203.409 &     19.2  & clear &  (27) \\
4204.432 &     18.7  & clear &  (25) \\
4205.358 &     17.6  & clear &  (24) \\
4205.417 &     17.7  & clear &  (23) \\
4205.489 &     17.7  & clear &  (23) \\
4206.361 &     18.1  & clear &  (27) \\
4206.402 &     18.1  & clear &  (23) \\
4206.410 &     18.4  & clear &  (28) \\
4207.375 &     18.3  & clear &  (24) \\
4207.417 &     18.4  & clear &  (23) \\
4208.330 &     18.5  & clear &  (28) \\
4208.353 &     18.5  & clear &  (26) \\
4208.409 &     18.3  & clear &  (30) \\
4209.394 &     18.6  & clear &  (23) \\
4210.339 &     18.9  & clear &  (26) \\
4211.409 &     18.9  & clear &  (23) \\
4212.391 &     19.0  & clear &  (29) \\
4213.389 &     19.0  & clear &  (23) \\
4216.391 &     19.6  & clear &  (27) \\
4218.377 &     20.1  & clear &  (23) \\
\end{longtable}
\normalsize

\scriptsize
\begin{longtable}[!hb]{llrrrr}
\caption{Photometry of nova candidates discovered by Neill\&Shara (2004).
Observers, observatories, telescopes and CCDs for measurements are coded in column comment according to Table~\ref{tab1}.
See caption of Table~\ref{tab1} for explanation of magnitude format.
Designation of nova candidates used here is the same as in Neill\&Shara (2004).}\label{tab5} \\
\hline\hline
JD& Mag & Band & Comment \\
(2\,450\,000+) \\
\hline
\endfirsthead
\caption{continued.}\\
\hline\hline
JD& Mag & Band & Comment \\
(2\,450\,000+) \\
\hline
\endhead
\hline
\endfoot
\noalign{\smallskip}
\multicolumn{6}{l}{Nova 1} \\
\noalign{\smallskip}
1912.626 & [21.5 & Ha & (2) \\
2632.684 & 21.24 &  r & (3) \\
2632.721 & 19.23 & Ha & (3) \\
2769.383 & [21.5 &  r & (7) \\
2769.388 & 20.77 & Ha & (7) \\
2781.432 & 20.8  & Ha & (4) \\
2781.436 & [21.7 &  r & (4) \\
2783.455 & 21.1  & Ha & (4) \\
2783.460 & [21.5 &  r & (4) \\
2785.450 & [21.0 & Ha & (4) \\
2794.391 & 20.9  & Ha & (5) \\
2794.397 & [21.3 &  r & (5) \\
2795.394 & 20.9: & Ha & (5) \\
2796.395 & 21.1  & Ha & (5) \\
2796.400 & [22.2 &  r & (5) \\
2800.384 & 21.2  & Ha & (6) \\
\noalign{\smallskip}
\multicolumn{6}{l}{Nova 2} \\
\noalign{\smallskip}
1912.626 & [21.5 & Ha & (2) \\
2632.684 & 20.85 &  r & (3) \\
2632.721 & 19.80 & Ha & (3) \\
2769.383 & [21.9 &  r & (7) \\
2769.388 & 21.4  & Ha & (7) \\
2781.432 & [21.6 & Ha & (4) \\
2781.436 & [22.4 &  r & (4) \\
2783.455 & 21.6  & Ha & (4) \\
2783.460 & [21.5 &  r & (4) \\
2796.395 & 21.8  & Ha & (5) \\
2800.384 & 21.4  & Ha & (6) \\
3039.528 & [22.3 &  V & (7) \\
3039.540 & [21.1 & Ha & (7) \\
3792.677 & [20.8 & Ha & (1) \\
3793.869 & [21.0 & Ha & (1) \\
\noalign{\smallskip}
\multicolumn{6}{l}{Nova 4} \\
\noalign{\smallskip}
1912.626 & [21.5 & Ha & (2) \\
2632.684 & [21.9 &  r & (3) \\
2632.721 & 19.09 & Ha & (3) \\
2781.432 & [21.6 & Ha & (4) \\
3039.528 & [22.3 &  V & (7) \\
3039.540 & [21.1 & Ha & (7) \\
3792.677 & [20.8 & Ha & (1) \\
3793.869 & [21.0 & Ha & (1) \\
\noalign{\smallskip}
\multicolumn{6}{l}{Nova 5} \\
\noalign{\smallskip}
1912.626 & [21.5 & Ha & (2) \\
2632.684 & [22.7 &  r & (3) \\
2632.721 & [21.2 & Ha & (3) \\
2769.383 & [21.7 &  r & (7) \\
2769.388 & 20.9  & Ha & (7) \\
2781.432 & 21.5  & Ha & (4) \\
2781.436 & [22.4 &  r & (4) \\
2783.455 & 21.4  & Ha & (4) \\
2783.460 & [21.5 &  r & (4) \\
2796.395 & 21.4  & Ha & (5) \\
2800.384 & 21.1  & Ha & (6) \\
3039.528 & [22.3 &  V & (7) \\
3039.540 & [21.1 & Ha & (7) \\
3792.677 & [20.8 & Ha & (1) \\
3793.869 & [21.0 & Ha & (1) \\
\noalign{\smallskip}
\multicolumn{6}{l}{Nova 7} \\
\noalign{\smallskip}
1912.626 & [21.5 & Ha & (2) \\
2632.684 & [22.7 &  r & (3) \\
2632.721 & [21.2 & Ha & (3) \\
2769.383 & [21.8 &  r & (7) \\
2769.388 & 21.0  & Ha & (7) \\
2781.432 & [21.6 & Ha & (4) \\
2781.436 & [22.4 &  r & (4) \\
2783.455 & 21.4  & Ha & (4) \\
2783.460 & [21.5 &  r & (4) \\
2796.395 & 21.5  & Ha & (5) \\
2800.384 & 21.5: & Ha & (6) \\
3039.528 & [22.3 &  V & (7) \\
3039.540 & [21.1 & Ha & (7) \\
3792.677 & [20.8 & Ha & (1) \\
3793.869 & [21.0 & Ha & (1) \\
\noalign{\smallskip}
\multicolumn{6}{l}{Nova 9} \\
\noalign{\smallskip}
1912.626 & [21.5 & Ha & (2) \\
2632.684 & [22.7 &  r & (3) \\
2632.721 & [21.2 & Ha & (3) \\
2769.383 & [21.8 &  r & (7) \\
2769.388 & 21.2  & Ha & (7) \\
2781.432 & [21.6 & Ha & (4) \\
2781.436 & [22.4 &  r & (4) \\
2783.455 & 21.6  & Ha & (4) \\
2783.460 & [21.5 &  r & (4) \\
2796.395 & [22.0 & Ha & (5) \\
2800.384 & [21.5 & Ha & (6) \\
3039.528 & [22.3 &  V & (7) \\
3039.540 & [21.1 & Ha & (7) \\
3792.677 & [20.8 & Ha & (1) \\
3793.869 & [21.0 & Ha & (1) \\
\noalign{\smallskip}
\multicolumn{6}{l}{Nova 10} \\
\noalign{\smallskip}
1912.626 & [21.5 & Ha & (2) \\
2632.684 & [22.7 &  r & (3) \\
2632.721 & [21.2 & Ha & (3) \\
2769.383 & [21.7 &  r & (7) \\
2769.388 & 20.24 & Ha & (7) \\
2781.432 & 20.6  & Ha & (4) \\
2781.436 & [22.4 &  r & (4) \\
2783.455 & 20.76 & Ha & (4) \\
2783.460 & [21.5 &  r & (4) \\
2785.450 & 20.6  & Ha & (4) \\
2794.391 & 21.0  & Ha & (5) \\
2794.397 & [21.3 &  r & (5) \\
2795.394 & 21.1  & Ha & (5) \\
2796.395 & 21.07 & Ha & (5) \\
2796.400 & [22.2 &  r & (5) \\
2800.384 & 20.9  & Ha & (6) \\
3039.528 & [22.3 &  V & (7) \\
3039.540 & [21.1 & Ha & (7) \\
3792.677 & [20.8 & Ha & (1) \\
3793.869 & [21.0 & Ha & (1) \\
\noalign{\smallskip}
\multicolumn{6}{l}{Nova 11} \\
\noalign{\smallskip}
1912.626 & [21.5 & Ha & (2) \\
2632.684 & [22.7 &  r & (3) \\
2632.721 & [21.2 & Ha & (3) \\
2769.383 & 20.60 &  r & (7) \\
2769.388 & 19.02 & Ha & (7) \\
2781.432 & 19.33 & Ha & (4) \\
2781.436 & 21.2  &  r & (4) \\
2783.455 & 19.32 & Ha & (4) \\
2783.460 & 20.95 &  r & (4) \\
2785.450 & 19.60 & Ha & (4) \\
2794.391 & 19.89 & Ha & (5) \\
2794.397 & [21.3 &  r & (5) \\
2795.394 & 20.12 & Ha & (5) \\
2796.395 & 19.97 & Ha & (5) \\
2796.400 & 21.6  &  r & (5) \\
2800.384 & 20.17 & Ha & (6) \\
3039.528 & [22.3 &  V & (7) \\
3039.540 & [21.1 & Ha & (7) \\
3792.677 & [20.8 & Ha & (1) \\
3793.869 & [21.0 & Ha & (1) \\
\noalign{\smallskip}
\multicolumn{6}{l}{Nova 12} \\
\noalign{\smallskip}
1912.626 & [21.5 & Ha & (2) \\
2632.684 & [22.7 &  r & (3) \\
2632.721 & [21.2 & Ha & (3) \\
2769.388 & [21.7 & Ha & (7) \\
2781.432 & 21.4  & Ha & (4) \\
2781.436 & 21.3  &  r & (4) \\
2783.455 & 21.0  & Ha & (4) \\
2783.460 & 20.46 &  r & (4) \\
2785.450 & 20.7  & Ha & (4) \\
2794.391 & 20.37 & Ha & (5) \\
2794.397 & 21.3: &  r & (5) \\
2795.394 & 20.32 & Ha & (5) \\
2796.395 & 20.16 & Ha & (5) \\
2796.400 & 21.9  &  r & (5) \\
2800.384 & 20.13 & Ha & (6) \\
3039.528 & [22.3 &  V & (7) \\
3039.540 & [21.1 & Ha & (7) \\
3792.677 & [20.8 & Ha & (1) \\
3793.869 & [21.0 & Ha & (1) \\
\end{longtable}
\normalsize

\end{document}